\documentclass{article}
\usepackage{ijcai24}

\usepackage{times}
\usepackage{soul}
\usepackage{url}
\usepackage[hidelinks]{hyperref}
\usepackage[utf8]{inputenc}
\usepackage[small]{caption}
\usepackage{graphicx}
\usepackage{amsmath}
\usepackage{amsthm}
\usepackage{booktabs}
\usepackage{algorithm}
\usepackage{algorithmic}
\usepackage[switch]{lineno}
\usepackage{multirow}
\begin{document}

\title{MuChin: A Chinese Colloquial Description Benchmark for Evaluating Language Models in the Field of Music}
% % Single author syntax
% \author{
%     Author Name
%     \affiliations
%     Affiliation
%     \emails
%     email@example.com
% }

% Multiple author syntax (remove the single-author syntax above and the \iffalse ... \fi here)
% \iffalse
\author{
Zihao Wang$^{1,2}$
\and
Shuyu Li$^{2}$\and
Tao Zhang$^2$\and
Qi Wang$^2$\and
Pengfei Yu$^2$\and
Jinyang Luo$^2$\and
Yan Liu$^2$\and
Ming Xi$^2$\And
Kejun Zhang$^{1}$\footnote{Corresponding Author}\\
\affiliations
$^1$Zhejiang University,\  $^2$DuiNiuTanQin Co., Ltd. \\
%$^3$Innovation Center of Yangtze River Delta,\  $^4$Xi’an Jiaotong University\\
% $^5$China University of Geosciences, Wuhan\\
\emails
carlwang@zju.edu.cn
\{lsyxary, zhangtao8, duoluo7161, cgoxopx,  rockyoungljy\}@gmail.com,
liuyan@liao.com,
ximing88@gmail.com,
zhangkejun@zju.edu.cn
}
% \fi

\maketitle

\begin{abstract}
    The rapidly evolving multimodal Large Language Models (LLMs) urgently require new benchmarks to uniformly evaluate their performance on understanding and textually describing music. However, due to semantic gaps between Music Information Retrieval (MIR) algorithms and human understanding, discrepancies between professionals and the public, and low precision of annotations, existing music description datasets cannot serve as benchmarks. To this end, we present MuChin, the first open-source music description benchmark in Chinese colloquial language, designed to evaluate the performance of multimodal LLMs in understanding and describing music. We established the Caichong Music Annotation Platform (CaiMAP) that employs an innovative multi-person, multi-stage assurance method, and recruited both amateurs and professionals to ensure the precision of annotations and alignment with popular semantics. Utilizing this method, we built a dataset with multi-dimensional, high-precision music annotations, the Caichong Music Dataset (CaiMD), and carefully selected 1,000 high-quality entries to serve as the test set for MuChin. Based on MuChin, we analyzed the discrepancies between professionals and amateurs in terms of music description, and empirically demonstrated the effectiveness of annotated data for fine-tuning LLMs. Ultimately, we employed MuChin to evaluate existing music understanding models on their ability to provide colloquial descriptions of music. All data related to the benchmark and the code for scoring have been open-sourced\footnote{\url{https://github.com/CarlWangChina/MuChin/}}.

\end{abstract}

\section{Introduction}
\label{sec:intro}
As Large Language Models (LLMs) have rapidly advanced, a multitude of LLMs have achieved notable results across various domains~\cite{zhao2023survey} and require comprehensive evaluation across benchmarks in different fields~\cite{liang2022holistic,huang2023ceval,chang2023survey}. Thus, the advancement of LLMs and multimodal technologies necessitates the establishment of benchmarks within the field of music for a unified evaluation. Although benchmarks currently exist for evaluating music understanding models, such as MARBLE~\cite{yuan2023marble}, which utilizes accuracy on downstream Music Information Retrieval (MIR) tasks as its metric, this does not comprehensively evaluate the capabilities of multimodal large language models.

\begin{figure*}[htp]
    \centering
    % \hspace*{-\dimexpr(\textwidth-1.1\columnwidth)/2\relax}
    \includegraphics[width=1.0\textwidth]{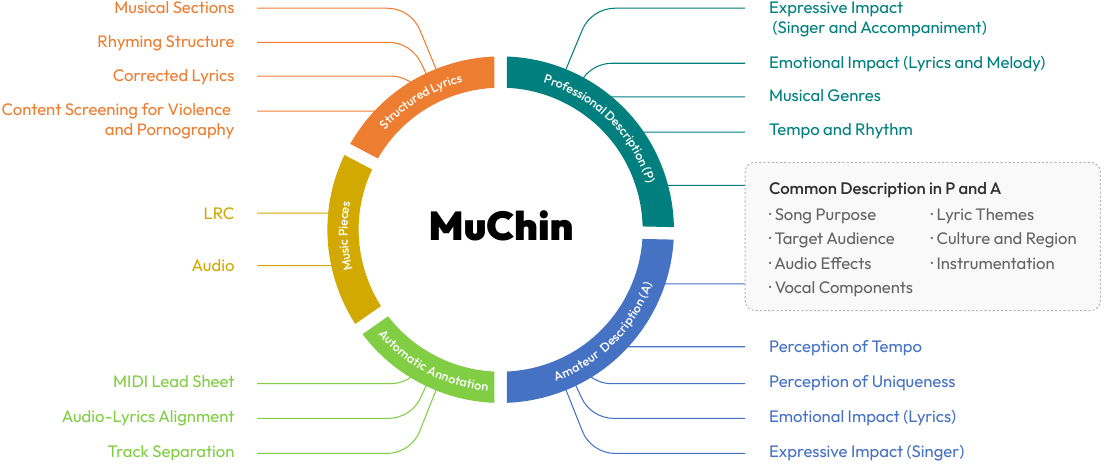}
    \caption{An overview of the MuChin benchmark. The Chinese Colloquial Descriptions consist of Description(A) and Common Description(P \& A) annotated by amateur annotators. In addition, we recruit professional annotators to label Description(P), Musical Sections, and Rhyming Structures of the lyrics. And machine-annotated information such as MIDI is also incorporated. These enable MuChin to adapt to a wider range of benchmark tasks.}
          
    % “muchin benchmark的概述图. Chinese Colloquial Description是由业余标注员标注的描述(A)和共同描述(P and A)组成. 此外我们还招募专业标注员标注了描述(P)、歌词的乐段和押韵结构, 并且用机器标注了MIDI等信息, 从而让Muchin能适应更多的benchmark tasks”
    \label{fig:MuChin}
    \vspace{-0.3cm}
\end{figure*}

Music description plays a crucial role in both music understanding~\cite{manco2021muscaps,gardner2023llark} and text-controlled music generation~\cite{agostinelli2023musiclm,copet2023musicgen}. However, there is currently a lack of benchmarks specifically for colloquial music description, which is why we introduce MuChin, the first open-source benchmark for Chinese colloquial music description, with details provided in \textbf{Figure~\ref{fig:MuChin}}.

As models for music understanding~\cite{jukemir,li2023mert} and music generation~\cite{zhang2023sdmuse,wang2023remast} have evolved, numerous datasets have been proposed, including those derived from Music Information Retrieval (MIR) algorithms or LLMs~\cite{bertin2011million,wang2020pop909,lu2023musecoco,huang2023noise2music,melechovsky2023mustango} as well as manually annotated datasets~\cite{yang2017midinet,bogdanov2019mtg,schneider2023mo,zhu2023ernie,wang2022songdriver,agostinelli2023musiclm}. However, these datasets present certain issues that prevent them from serving as comprehensive benchmarks to thoroughly evaluate models' performance in understanding and describing music. Firstly, there is a considerable semantic gap between datasets obtained through algorithms and complex human descriptions. Secondly, current datasets annotated manually are confined to expert annotations and limited descriptive scopes, which significantly diverge from the descriptions provided by the general public
% existing manual annotated datasets are limited to professional annotations and narrow descriptive dimensions, which still differ significantly from the descriptions given by the public
~\cite{amer2013older,mikutta2014professional}. And a detailed discussion will be presented in \textbf{Section~\ref{sec:p_vs_a}}. Thirdly, due to limitations in algorithms' performance, datasets generated by MIR cannot achieve complete accuracy, and existing manually annotated datasets, where each entry is annotated by only one person, can also be prone to inaccuracies caused by human errors or biases.

% To address these issues, it is necessary to recruit both professionals and amateurs to annotate music and produce, respectively, professional music descriptions rich in technical musical terminology, and amateur music descriptions that align with the public's colloquial descriptions. Additionally, we implemented a complex multi-person, multi-stage assurance mechanism to ensure the accuracy of the annotations.

To tackle these challenges, we need to engage both professionals and amateurs in annotating music. This approach will yield two distinct types of music descriptions: one, from professionals, will be rich in technical musical terms, while the other, from amateurs, will resonate with the general public's everyday language. Furthermore, we have introduced a sophisticated, multi-tiered quality assurance process involving multiple individuals at various phases to guarantee the precision of these annotations.  %TZ

% Based on this design, we developed a platform that suggests popular music descriptive terms from the web or professional musical terms, based on the words entered by the annotator, allowing annotators to quickly find the desired descriptions. Moreover, the platform's backend is capable of ensuring the accuracy of annotations by utilizing a multi-person, multi-stage assurance mechanism. This method can improve the efficiency, accuracy, and consistency of annotators' descriptions, while also aligning with the public since the descriptive terms are sourced from the web.

Building on this design, we created a platform that recommends widely-used music descriptors from the internet or specialized terms from the music industry, depending on the input from the annotator. This feature enables annotators to swiftly locate the precise descriptions they need. Additionally, the platform's backend employs a multi-layered, multi-person quality assurance process to verify the precision of the annotations. This approach enhances the efficiency, precision, and uniformity of the annotators' descriptions and ensures relevance to the general public by sourcing descriptive terms directly from the web. %TZ

% Utilizing this platform, we built a multi-dimensional, high-precision, user-aligned dataset with over 100k entries, referred to as the Caichong Music Dataset (CaiMD). From this, we carefully selected 1k high-quality entries as test set and established a benchmark for evaluating language models on generation and understanding tasks in music. Considering the precision of these annotated data, they can also be utilized for fine-tuning pre-trained LLMs on various music downstream tasks. Therefore, we extracted additional 9k entries of data from CaiMD for training and then fine-tuned an LLM based on Qwen~\cite{qwen}, demonstrating the effectiveness of our data for fine-tuning LLMs and the effect of fine-tuning techniques in the field of music.

With this platform, we have developed a comprehensive, highly accurate, and public-aligned dataset, the Caichong Music Dataset (CaiMD).
% From this extensive collection, we meticulously selected 1,000 high-quality entries 
From this extensive collection, we meticulously selected 1,000 high-quality entries to serve as a test set, thereby establishing a benchmark for evaluating language models' capabilities in both generating and understanding music-related tasks. Given the precision of these annotated entries, they are also exceptionally suited for fine-tuning pre-trained large language models (LLMs) for a variety of music-related downstream tasks. 
% To this end, we extracted an additional 9,000 entries from CaiMD and subsequently fine-tuned a LLM based on the Qwen~\cite{qwen} model. This process demonstrated the effectiveness of our data on fine-tuning LLMs. %TZ
To illustrate this point, we have fine-tuned an LLM with an additional dataset, thereby demonstrating its efficacy.

% MuChin provides a new perspective on the performance of language models in the field of music, requiring models not only to extract basic attributes from music and describe it from a professional standpoint but also to align with the public. The models need to understand descriptions provided by amateur and be capable of describing music in a manner that is accessible to the public.

MuChin provides a new perspective on the performance of language models in the field of music, requiring the model not only to extract basic attributes from music and describe it from a professional point of view, but also to be able to align with the musical feelings of public users, and describe music in a popular way. %TZ

Our Contributions are:
\begin{enumerate}
    \item We proposed and open-sourced MuChin: the first Chinese colloquial music description benchmark designed to more comprehensively assess the capabilities of multimodal LLMs in the field of music. Utilizing this benchmark, we evaluated the performance of existing music understanding models in terms of their ability to describe music colloquially, as well as the proficiency of current LLMs in generating structured lyrics.
    
    % \item We developed the Caichong Music Annotation Platform (CaiMAP) with a multi-person, multi-stage assurance method to ensure the accuracy and consistency of annotations, ultimately achieving the goal of efficiently annotating both professional and colloquial music descriptions, as well as musical sections and rhyming.

    \item We created the Caichong Music Annotation Platform (CaiMAP), implementing a multi-person, multi-stage quality assurance process to guarantee the precision and uniformity of annotations. This approach successfully facilitates efficient annotation of both professional and colloquial music descriptions, including musical sections and rhymes. %TZ 
    
    \item We built the Caichong Music Dataset (CaiMD): a dataset that is multi-dimensional and high-precision, aligned with the public. It contains music annotations encompassing information on both professional and colloquial descriptions. Through empirical studies, we demonstrated the effectiveness of the CaiMD on fine-tuning LLMs. Furthermore, we analyzed and verified the discrepancies between professionals and amateurs in terms of music understanding and description.
\end{enumerate}

% \vspace{0.30cm}

\section{Establishment of MuChin Benchmark}

% To address the gap in Chinese colloquial benchmarks for language models in the field of music, we collected and built an annotated dataset, upon which we established the MuChin benchmark. In this section, we will provide an introduction to the establishment of MuChin.

To bridge the gap in benchmarks for language models within the domain of music, specifically targeting Chinese colloquial expressions, we curated and constructed an annotated dataset. This effort led to the creation of the MuChin benchmark.  %TZ

\vspace{0.30cm}

\subsection{Benchmark Tasks}
\label{sec:benchmark_tasks}

% For the purpose of evaluating LLMs in various dimensions, we incorporate various tasks as follows in our dataset, and thus establishing MuChin based on these tasks.

To assess LLMs across multiple dimensions, we included a variety of tasks in our dataset, leading to the creation of MuChin, which is based on the following tasks. %TZ

\label{sec:tasks}

\subsubsection{Textual Description Task}

% Textual descriptions of music involve multi-dimensional characterizations such as auditory perception, emotions, and categorization of music, which include tagging and integrated textual descriptions. This annotated data can be utilized to establish a benchmark for evaluating the music understanding capabilities of multimodal LLMs on tasks like music emotion recognition and music classification. Additionally, it serves to evaluate the capacity of LLMs to process descriptive texts about music. Furthermore, this data can also be employed for fine-tuning LLMs with music-related text, thereby enhancing their performance on music-related tasks.

Textual descriptions of music involve multi-dimensional representations, including auditory perception, emotions, and music classification.  Annotators are required to label and write textual descriptions.  Such annotated data sets the stage for benchmarking the ability of multimodal LLMs in understanding music, particularly in tasks like music emotion recognition and classification.  Moreover, this data facilitates the evaluation of LLMs' capacity in processing descriptive music texts. Additionally, it can be used to fine-tune LLMs with music-related content. %TZ

When annotating textual descriptions, annotators are required to describe music from various aspects, as shown in \textbf{Figure~\ref{fig:MuChin}}. To enhance the efficiency, precision, and consistency of annotations, and to align with the public, we built lexicons of music descriptive terms, including a popular term lexicon and a professional term lexicon. The former consists of popular music descriptive terms collected from the internet, while the latter contains keywords extracted from the descriptions of the open-source text-music dataset MusicCaps~\cite{agostinelli2023musiclm}. % Annotators can select suitable terms from the pre-built lexicon or choose to supplement the descriptions with their own input if they are dissatisfied with the terms in the lexicon.
Annotators have the option to choose appropriate terms from an existing lexicon or, if they find the terms in the lexicon unsatisfactory, they can enhance the descriptions with their own contributions. %TZ

% As mentioned in Section~\ref{sec:intro}, in the current manual annotation process, annotators struggle to find accurate descriptive words to describe music, which affects the efficiency and consistency of the annotations. To address this issue, we devise a solution involving recommended words in conjunction with user-supplied input. Specifically, we pre-built a lexicon of music descriptive terms based on popular terms on the web and specialized musical terminology. Annotators can select suitable terms from the pre-built library during annotation, or use keyword searches to find descriptive tags that meet their requirements. Additionally, if annotators are dissatisfied with the terms in the library, they can also choose to supplement the descriptions with their own input.

% \begin{table}
%     \centering
%     \begin{tabular}{lll}
%         \hline
%         Scenario  & $\delta$ & Runtime \\
%         \hline
%         Paris     & 0.1s     & 13.65ms \\
%         Paris     & 0.2s     & 0.01ms  \\
%         New York  & 0.1s     & 92.50ms \\
%         Singapore & 0.1s     & 33.33ms \\
%         Singapore & 0.2s     & 23.01ms \\
%         \hline
%     \end{tabular}
%     \caption{Latex default table}
%     \label{tab:plain}
% \end{table}

\subsubsection{Lyric Generation Task}

% Lyric generation represents a significant application of LLMs in the field of music, demanding that LLMs possess a deep understanding of musical forms to generate structured lyrics. Therefore, we build our dataset with lyric structure information and then establish a benchmark for evaluating the performance of LLMs in structured lyric generation. This means we need to clearly demarcate each section of the lyrics, resulting in the generation of lyrics with distinct musical sections.

Lyric generation stands as a notable use case for LLMs within the music industry, requiring LLMs to have a profound comprehension of musical structures in order to produce well-organized lyrics. To facilitate this, we construct our dataset to include information on lyric structure, thereby setting a benchmark for assessing LLMs' proficiency in generating lyrics with clear structural distinctions. This involves meticulously defining each section of the lyrics. %TZ

Additionally, the ability of LLMs to generate lyrics that align with the theme and rhyme is also crucial. Thus, annotators are required to annotate the main themes and rhymes, as well as to correct any textual errors within the lyrics.

\subsubsection{Tasks with Automatic Annotation}

Tasks with automatic annotation are discussed in \textbf{Appendix~\ref{app:auto_anno}.}

\subsection{Preparation and Settings}
\label{sec:pre_n_set}

For the benchmark tasks delineated in \textbf{Section~\ref{sec:benchmark_tasks}}, it is essential to annotate the data across the corresponding dimensions. % Consequently, in this section, we will engage in data preprocessing and individual recruitment and training, striving to achieve comprehensive and high-precision annotations.
Therefore, in this section, we will undertake data preprocessing, along with the recruitment and training of individuals, aiming to secure thorough and high-precision annotations. %TZ

\subsubsection{Data Preprocessing}

Data preprocessing, including \textbf{music genre clustering}, \textbf{track separation}, \textbf{audio-lyrics alignment}, and \textbf{automatic pre-annotation} is provided in \textbf{Appendix~\ref{app:data_pro}}.

\subsubsection{Recruitment and Training of Individuals}

% In order to annotate music with both amateur and professional descriptions, we need to recruit amateur music enthusiasts to annotate music with popular descriptive terms, as well as professionals, including students majoring in music and music practitioners, to act as specialized annotators and quality assurance inspectors. Guided by this concept, we recruit 213 individuals familiar with Chinese through campus and public recruitment initiatives, including 109 amateur music enthusiasts and 104 professionals. The participants include 144 males and 69 females, with an age range of 19 to 35 years old. We divide the individuals into four groups, with each group's specific tasks as follows:

To annotate music using both amateur and professional descriptions, it is necessary to engage amateur music enthusiasts for annotating music with popular terms, and professionals -- including music students and practitioners -- as specialized annotators and quality assurance inspectors.  Following this approach, we have recruited 213 individuals familiar with Chinese music through campus and public recruitment efforts. This group includes 109 amateur music enthusiasts and 104 professionals, consisting of 144 males and 69 females, with ages ranging from 19 to 35 years. We have organized these participants into four groups, each assigned specific tasks as follows:  %TZ

\begin{itemize}
    \item \textbf{Professional Group.} Annotate structures, rhymes and provide professional descriptions.
    \item \textbf{Amateur Group.} Provide colloquial descriptions.
    \item \textbf{Inspector Group.} Evaluate structure annotations, and score music descriptions.
    \item \textbf{Administrator.} Address and provide feedback on inquiries from various groups, and conduct random spot-checks of the groups' outcomes.
\end{itemize}

The \textbf{grouping} and \textbf{training} \textbf{method} for each group of individuals are detailed in the \textbf{Appendix~\ref{app:individual}}.

\subsection{Annotation and Assurance Pipeline}

%Following the preparation is the manual annotation period. On the one hand, to enhance the quality of annotations and ensure their accuracy as much as possible, and on the other hand, to objectively evaluate the annotators' performance, we devised an innovative multi-person, multi-stage assurance method. Based on this method, we developed the \textbf{Caichong Music Annotation Platform (CaiMAP)}, which is introduced in \textbf{Appendix~\ref{app:caimap}.} The specific annotation pipeline is shown as Figure~\ref{fig:pipeline} and will be introduced in this section.

\begin{figure*}[htb]
    \centering
    \includegraphics[width=\textwidth]{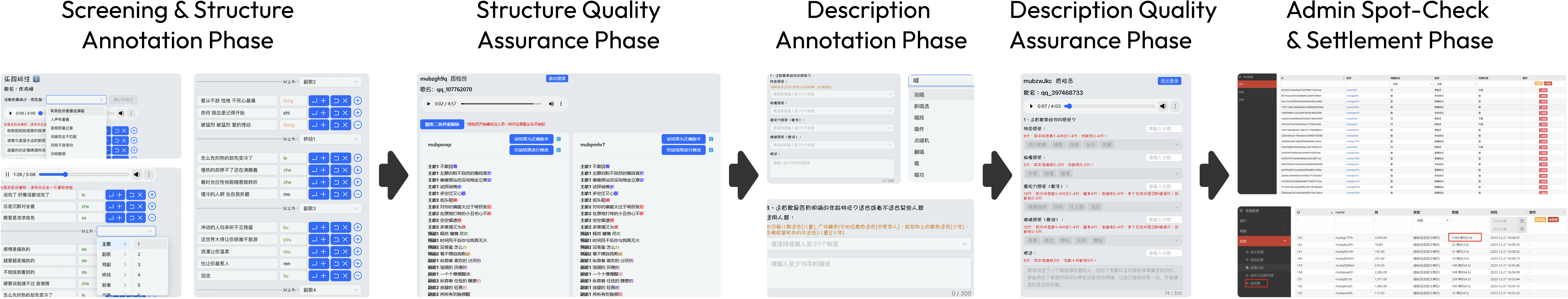}
    \caption{Pipeline of data annotation and assurance. Each annotated data undergoes 5 complex phases to ensure the accuracy. The figure shows the actual screenshots of the pages for each phase. For \textbf{software development} and \textbf{operation} details please refer to \textbf{Appendix \ref{app:caimap}}.}
    
    \label{fig:pipeline}
    \vspace{-0.3cm}
\end{figure*}

The subsequent phase involves annotation.  We have devised an innovative multi-person, multi-stage assurance method aimed at improving quality of annotations and maximizing their accuracy.  Additionally, this method serves to objectively evaluate the performance of annotators. Based on this method, we developed the \textbf{Caichong Music Annotation Platform (CaiMAP)}, which is introduced in \textbf{Appendix~\ref{app:caimap}.}  The specific annotation pipeline is shown as Figure~\ref{fig:pipeline} and will be introduced in this section.  %TZ

% and most are either current university students or individuals who have graduated within the past five years. Among these participants, the professionals consist of 68 males and 36 females, all of whom are either current music majors or graduates within the past five years, each with over three years of experience in music-related coursework. The amateur enthusiasts comprise 76 males and 33 females, none of whom have received formal music education.

\subsubsection{Screening \& Structure Annotation Phase}

In the screening phase, annotators are required to screen the data carefully. Music pieces with poor audio quality or content involving pornography or violence that are unsuitable for the dataset should be skipped.  %TZ

% After selecting a song, the platform automatically begins to play the music, and annotators are not allowed to adjust the progress bar or playback speed until the entire song has finished playing. The platform presents the complete lyrics sentence by sentence, and annotators are required to insert musical section tags between the lyrics. Regarding rhyming, annotators are required to check the accuracy of the pre-annotated phonemes and rhymes for each line. If any inaccuracies are found, they should provide their own annotations.

In the structure annotation phase, the platform presents the complete lyrics sentence by sentence, and annotators are required to insert musical section tags between the lyrics. Annotators are also required to check the accuracy of the pre-annotated phonemes and rhymes for each line. If any inaccuracies are found, they should provide their own annotations.  %TZ

\subsubsection{Structure Quality Assurance Phase}

% After a piece of data being annotated by two annotators, the platform automatically determines whether their annotations match. If they are identical, the platform directly incorporates the annotated data into the dataset, awaiting the next phase of annotation. If there is a discrepancy, the annotations from both annotators are submitted and presented to a quality assurance inspector, who compares them and identifies the correct annotations, making modifications to certain data or re-annotating it entirely. The annotations submitted by the inspector are then included in the structure annotation dataset as the correct annotations. Meanwhile, the platform records the accuracy rate of both annotators in the backend, and subsequent screening is conducted according to the settings in Section~\ref{sec:pre_n_set}.

% To ensure the accuracy of the annotations, we designed an assurance mechanism. Each data should be annotated by two different annotators, after which the platform automatically determines whether the annotations match. If they are the same, the platform will directly incorporate this data into the dataset for the next phase. In cases of discrepancies, both sets of annotations will be submitted to a quality assurance inspector to determine which is the correct result, or to submit a correct annotation independently as the result.

To ensure the accuracy of the annotations, we implemented a quality assurance mechanism. Each piece of data undergoes annotation by two separate annotators. Subsequently, the platform autonomously verifies the congruence of the annotations. If they align, the platform seamlessly integrates the data into the dataset for the subsequent phase. In instances of disparities, both sets of annotations are referred to a quality assurance inspector for resolution. The inspector determines the correct annotation or submits an independent correction if necessary.  %TZ

\subsubsection{Description Annotation Phase}

Data that successfully clears the structure quality assurance phase becomes eligible for utilization in the music description phase. During this phase, to guarantee attentive listening and thoughtful music descriptions, annotators must listen to each music piece without interruption. Specifically, annotators are prohibited from writing any textual descriptions within the initial 30 seconds of the music piece.  Copy and paste content is also not allowed.  Additionally, limitations are imposed on the number of tags that can be entered and on the word count of user-generated entries.  %TZ

\subsubsection{Description Quality Assurance Phase}

% Each week, the platform randomly selects 20\% of the annotation results from each annotator and submits these to the inspectors for scoring. These scores are recorded in the platform backend, and subsequent screening is conducted according to the settings in Section~\ref{sec:pre_n_set}. Annotated data that pass the sampling quality assurance are incorporated into the dataset, while those that fail are entirely rejected.

% Given that music description annotation is subjective and difficult to evaluate, the platform randomly selects 20\% of the annotation results from each annotator and submits these to the quality assurance inspectors for scoring. These scores are recorded in the backend of the platform. Annotated data that pass the sampling quality assurance are incorporated into the dataset, while those that fail are entirely rejected.

Since music description annotation involves subjective judgments and is challenging to assess, the platform employs a randomized selection process, choosing 20\% of the annotation results from each annotator for submission to quality assurance inspectors for scoring. These scores are then logged in the platform's backend. Annotated data that successfully pass the sampling quality assurance are submitted into the dataset, whereas those that do not meet the standards are rejected. %TZ

\subsubsection{Admin Spot-Check \& Settlement Phase}

% During the annotation and quality assurance processes, administrators review the results of both annotators and inspectors. Platform administrators have access to viewing all submissions, including the annotation results of all annotators, the correct annotations submitted by all inspectors, and all scores given by inspectors to textual descriptions. Administrators can enter the account they wish to review in the administrator interface and retrieve all the aforementioned information related to that account. In this manner, administrators can spot-check the work of each user, including annotators and inspectors, to ensure the efficiency and accuracy of the annotation tasks. Moreover, users can report any platform bugs or issues they encounter while using the platform to the administrators. Upon receiving these reports, administrators either provide solutions or contact developers to resolve the issues.

% Administrators can review the work status of each group in real-time and settle payments based on the results of the quality assurance. Annotators with a high pass rate for their annotations will receive additional rewards, while those with a lower pass rate will face certain penalties, thereby encouraging them to annotate diligently.

Administrators can monitor the real-time progress of each group's work and make payments accordingly, depending on the outcomes of quality assurance checks. Annotators who consistently achieve high pass rates for their annotations will be rewarded additionally, whereas those with lower pass rates will incur penalties, thus motivating them to annotate diligently.  %TZ

To determine whether the inspectors are competent in their work, administrators also have the access to randomly selected samples of their work for secondary verification.

% Furthermore, administrators can track user behaviors, such as duration on the page and frequency of interactions, to gauge their level of attention. Warnings are issued to annotators whose performance appears to be perfunctory. Annotators who accumulate multiple warnings will no longer be eligible to participate in subsequent annotation tasks.

All the qualified annotated data are incorporated into the \textbf{CaiMD}. We provide the subsequent \textbf{data processing procedures}, \textbf{examples}, and an \textbf{overview} in \textbf{Appendix~\ref{app:caimd}}.

\section{Experiments}

% In this section, we will initially analyze the discrepancies between professionals and amateurs, thus emphasizing the significance of the alignment to the public. And then we will select several recent language models as baselines, involving both generative LLMs and music understanding models. Subsequently, we will evaluate their capability to understand music, understand music descriptions and complete downstream tasks. By implementing these experiments, we aim to evaluate the capability of the recent language models in the field of music, as well as provide a demonstration of our benchmark.

In this section, we will begin by examining the disparities between professionals and amateurs, thereby underscoring the importance of alignment with public perception. Following that, we will choose several recent language models as benchmarks, encompassing both generative language and music comprehension models. We will then assess their ability to comprehend music, understand musical descriptions, and perform downstream tasks. Through these experiments, our goal is to evaluate the effectiveness of recent language models in the realm of music and to demonstrate our benchmarking approach.  %TZ

\begin{figure*}
    \centering
    \includegraphics[width=\textwidth]{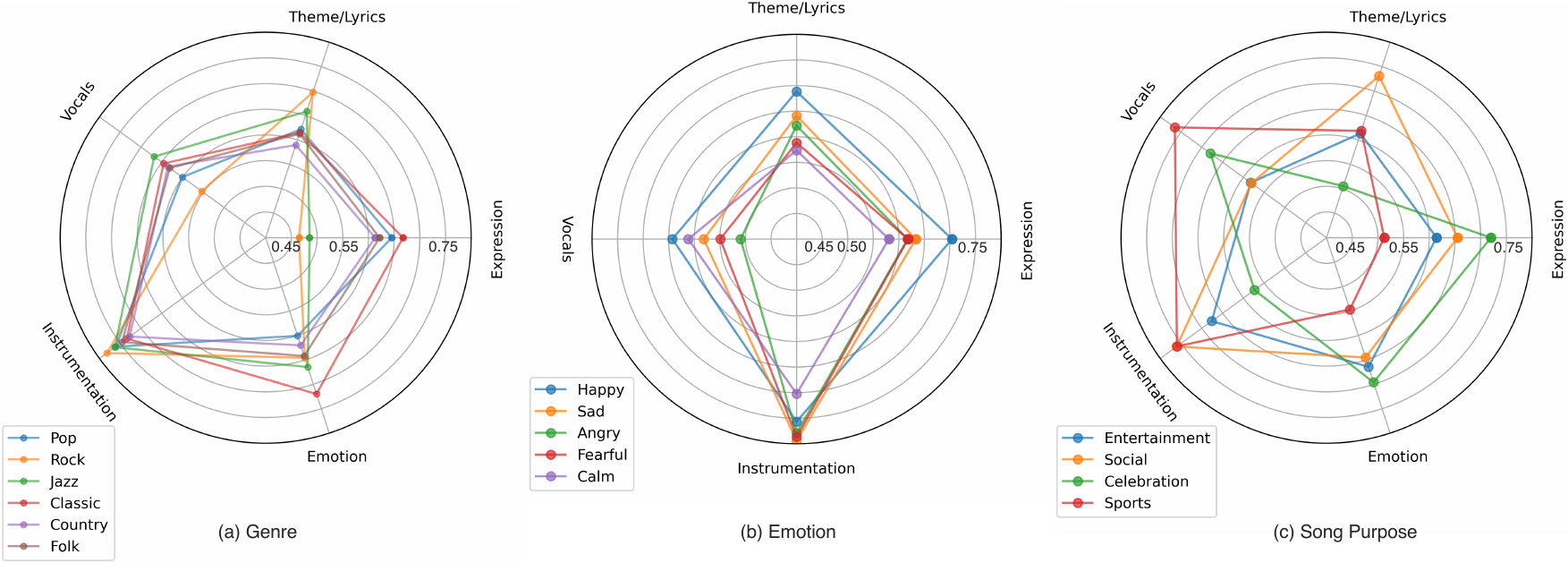}
   
    \caption{Semantic similarity scores between professionals and amateurs. When a specific type of music is selected, we calculate the similarity between the two groups in various dimensions, for which the \textbf{calculation method} is discussed in \textbf{Section~\ref{sec:und}}. As a \textbf{smaller} value signifies a\textbf{ larger discrepancy}, the experimental results in this figure reveal significant gaps between the two groups across several specific dimensions.}
    \label{fig:vs_result}
    \vspace{-0.25cm}
\end{figure*}

\subsection{Discrepancies Between Professionals and Amateurs}
\label{sec:p_vs_a}

% We discover in the CaiMD that there are significant discrepancies in the descriptions of certain musical attributes of the same music piece between amateur and professional groups. We believe this may be associated with the different levels of musical education the two groups have received. The professional group consists of individuals majoring in music-related programs, while the amateur group has not undergone formal music education and simply enjoys listening to music as a hobby. Consequently, based on the 10k entries of data we previously selected, we conduct a study on the cognitive biases between the two groups regarding various musical attributes, with the hope of inspiring research on the impact of education on human cognition.

% To demonstrate that there is indeed a significant gap between the understanding and description of music by professionals and amateurs, such that the descriptions by professionals cannot align with the public, we conducted an experiment to measure the discrepancies in how these two groups describe various musical attributes across different dimensions.

To illustrate the substantial disparity between the comprehension and description of music by professionals and amateurs, highlighting the inability of professional descriptions to resonate with the public, we conducted an experiment to gauge the differences in how these two groups articulate various musical attributes across various dimensions.  %TZ

\subsubsection{Analysis Metrics}

When a specific type of musical attributes is selected, we calculate the semantic similarity between professionals and amateurs across various dimensions, utilizing the \textbf{Semantic Similarity Score }metric which will be detailed in \textbf{Section~\ref{sec:und}}.

\subsubsection{Results}

The results of the discrepancies between professionals and amateurs across various dimensions are as \textbf{Figure~\ref{fig:vs_result}}.

From \textbf{Figure~\ref{fig:vs_result}(a)}, % it can be observed that there is little difference in the multidimensional descriptions of most music genres between the two groups. However, there are significant discrepancies in their perceptions of expression in Jazz and Rock. This suggests that there are considerable discrepancies in the understanding and descriptions of expression in progressive genres between professionals and amateurs.
it is evident that there is minimal variance in the multidimensional descriptions of most music genres between the two groups. However, notable disparities arise in their perception of expression in Jazz and Rock, implying significant differences in understanding and describing of expression within progressive genres between professionals and amateurs. %TZ

From \textbf{Figure~\ref{fig:vs_result}(b)}, % it can be seen that there is a greater discrepancy between professionals and amateurs in their interpretations of songs with calm and angry emotions compared to happy ones. This indicates that emotions also affect the understanding between the two groups.
a greater discrepancy between professionals and amateurs is apparent in their interpretations of music pieces evoking calm and angry emotions, in contrast to those evoking happiness. This underscores the impact of emotions on the comprehension divide between the two groups. %TZ

% From \textbf{Figure~\ref{fig:vs_result}(c)}, it can be observed that there are significant discrepancies in the distribution of the semantic similarity for different song purposes. This indicates that the two groups have varying dimensional discrepancies in their understanding of music with different purposes.

\textbf{Figure~\ref{fig:vs_result}(c)} reveals substantial disparities in the semantic similarity distribution across various song purposes. This discrepancy suggests that professionals and amateurs have distinct dimensional understandings of music tailored to different intents. %TZ

% Taking these into consideration, we believe that professionals and amateurs exhibit varying degrees of interpretative discrepancies across different dimensions for various types of music. Therefore, a comprehensive music description benchmark must be capable of encompassing both groups.
Considering these findings, it becomes evident that professionals and amateurs exhibit varying levels of interpretative disparities across diverse dimensions and music types. Therefore, a comprehensive music description benchmark should accommodate both groups' perspectives. %TZ

\subsection{Generative LLMs}

We utilize MuChin to evaluate existing LLMs in structured lyric generation, including Qwen~\cite{qwen}, Baichuan-2~\cite{baichuan2023baichuan2}, GLM-130B~\cite{zeng2022glm}, and GPT-4~\cite{achiam2023gpt}. % Additionally, considering that Qwen is primarily trained on Chinese corpus and performs well in the Chinese language environment, we also fine-tuned Qwen using 9,000 entries of the CaiMD dataset and then assess this fine-tuned Qwen model on MuChin to evaluate the effectiveness of our dataset in fine-tuning LLMs as well as the fine-tuned LLMs' ability to understand music descriptions and complete corresponding tasks.
Moreover, taking into account that Qwen is primarily trained on a Chinese corpus and excels in Chinese language environments, we further refined Qwen by fine-tuning it with another batch of data. Subsequently, we evaluated the performance of this fine-tuned Qwen model on MuChin to assess both the efficacy of the data in fine-tuning language and music models, as well as the fine-tuned model's proficiency in comprehending music descriptions and executing associated tasks. %TZ

\begin{table*}[htb!]
\centering
\resizebox{\textwidth}{!}{%
\renewcommand\arraystretch{1.3}
\begin{tabular}{rrrrrrr}

\toprule
\multirow{2}{*}{\textbf{Model}} & \multirow{2}{*}{\textbf{}} & \multirow{2}{*}{\textbf{GPT-4}} & \multirow{2}{*}{\textbf{GLM-4}} & \multirow{2}{*}{\textbf{Baichuan-2}} & \multicolumn{2}{c}{\textbf{Qwen}}\\  \cline{6-7} 
 &  &  &  &  &  \textbf{Base Model}  & \textbf{Fine-tuned}  \\  \hline 
 Parameter Size &  & 1800B & 130B & 53B & 14B & 14B \\
Overall Score&  & \underline{67.08(±6.23)} & 54.93(±16.46) & 49.19(±15.85) & 48.31(±13.39) & \textbf{85.24(±11.65)}\\
\hline
\multirow{4}{*}{Structure Similarity} & Song Level & 2.50(±1.16) & 2.29(±0.97) & 2.32(±0.99) & \underline{2.58(±1.51)} & \textbf{4.69(±2.38)}\\
 & Section Level & \underline{32.40(±0.41)} & 28.20(±6.75) & 28.83(±8.02) & 26.49(±4.92) & \textbf{32.14(±0.91)}\\
 & Phrase Level & \underline{15.52(±2.19)} & 12.93(±4.31) & 12.74(±4.36) & 11.59(±3.80) & \textbf{17.01(±0.80)} \\
 & Word Level & \underline{0.36(±0.79)} & 0.15(±0.39) & 0.01(±0.02) & 0.10(±0.23) & \textbf{9.12(±5.92)} \\ \hline
\multirow{2}{*}{Rhyming} & Fitting Accuracy & \underline{13.88(±3.05)} & 9.61(±5.17) & 4.84(±4.72) & 8.01(±4.36) & \textbf{16.30(±2.94)} \\
 & Proportion Reasonableness & \underline{2.40(±2.66)} & 1.74(±2.65) & 0.45(±1.96) & 1.29(±1.89) & \textbf{5.98(±4.03)} \\ \bottomrule
\end{tabular}%
}
\caption{Evaluation results of the selected LLMs on the benchmark of structured lyric generation. The results are\textbf{ calculated} by the \textbf{formula} detailed in \textbf{Appendix~\ref{app:eval_lyr}}. A larger value indicates a higher degree of similarity to the corresponding dimension of the actual lyrics, signifying better quality of the generated structured lyrics. For base models, the highest score in each dimension is underlined.}
\label{tab:llm_results}
\vspace{-0.2cm}
\end{table*}

\subsubsection{Evaluation Metrics}

% To evaluate the performance of LLMs, we input music description prompts into them, requesting them to output structured lyrics with musical sections and rhyming. While the content of the lyrics should exhibit subjective diversity, the structure of the lyrics is objective. Therefore, we focus primarily on the accuracy of the lyric structure rather than its content. We present an evaluation method that assesses the similarity between the model-generated lyrics and the ground truth across 6 dimensions as follows.

In assessing the performance of LLMs, we prompt them with music description inputs, asking for structured lyrics along with musical sections and rhymes. While the lyrical content should present subjective diversity, the structural integrity remains objective. Hence, our evaluation primarily centers on the accuracy of the lyric structure rather than its content. We introduce an evaluation method that measures the likeness between the model-generated lyrics and the ground truth across six dimensions outlined below. %TZ

\begin{itemize}
    \item \textbf{Song Level.} Song structure similarity measures the similarity between the generated lyrics and the ground truth in terms of overall structure.

    \item \textbf{Section Level.} Section structure similarity measures the similarity between the generated lyrics and the ground truth in terms of musical section labels, order, and the number of sections.

    \item \textbf{Phrase Level.} Phrase structure similarity measures the similarity in the number of phrases within each musical section compared to the ground truth.

    \item \textbf{Word Level.} Word structure similarity measures the similarity between the generated lyrics and the ground truth in terms of the number of words per corresponding phrase.

    \item \textbf{Rhyming Fitting Accuracy.} Rhyme fitting accuracy measures the degree to which the generated lyrics match the ground truth, in terms of end-of-line rhymes.

    \item \textbf{Rhyming Proportion Reasonableness.} To further measure the reasonableness of rhyming, we set an additional reward score based on the proportion of rhyming sentences within the overall lyrics, to evaluate the reasonableness of the rhyming proportion in the generated lyrics.
\end{itemize}

% The overall similarity is calculated using a weighted average with weights of 0.10, 0.325, 0.175, 0.20, and 0.20 respectively for the first five dimensions. And the rhyming proportion reasonableness is given an extra weight of 0.10.
The overall similarity is calculated by computing a weighted average, with weights of 0.10, 0.325, 0.175, 0.20, and 0.20 assigned respectively to the first five dimensions: song, section, phase, word, and rhyming fitting. Additionally, an extra weight of 0.10 is allocated to assess the reasonableness of rhyming proportions.

After comprehensive consideration, the Gestalt algorithm~\cite{ratcliff1988pattern}, which is a universal algorithm for string matching and similarity calculation, is suitable for our lyric evaluation task. Based on the Gestalt algorithm, we propose a scoring algorithm to assess the similarity between generated lyrics and actual lyrics.

The \textbf{calculation of the scores} of different dimensions is detailed in \textbf{Appendix~\ref{app:eval_lyr}}.

\subsubsection{Results}

\textbf{Table~\ref{tab:llm_results}} presents the similarity scores across various dimensions for structured lyrics generated by the selected LLMs in a one-shot scenario, utilizing music descriptions as provided prompts. Notably, all models achieve commendable results. We can observe that among the base models, the overall score increases with the expansion of parameter size. Thanks to its vast parameter size and extensive training data, GPT-4 significantly outperforms the other three models across most dimensions. However, the fine-tuned Qwen, despite having fewer parameters, notably surpasses the untuned base models in overall score and demonstrates a substantial lead in every dimension. This underscores the significant impact of fine-tuning in enhancing the model's capability to comprehend music descriptions and generate structured lyrics. It also suggests considerable potential for improvement in current LLMs within the field of music, emphasizing the importance of MuChin in advancing the development of Chinese LLMs in this domain. %TZ

% \begin{table}[htb]
%     \centering
%     \begin{tabular}{c|c}
          
%     \end{tabular}
%     \caption{The results have been calculated through cumulative multiplication and weighting according to Equation~\ref{eq:overall}, and then multiplied by 100 to serve as the scores.}
%     \label{tab:llm_results}
% \end{table}

\subsection{Music Understanding Models}
\label{sec:und}

Analogous to pre-trained language models in NLP, such as BERT~\cite{devlin2018bert}, a proficient pre-trained music understanding model should be able to effectively represent information across various dimensions within the music, allowing it to be extracted using a simple shallow neural network acting as a decoder. In our benchmark tailored for Chinese music description, we primarily evaluate the capabilities of music understanding models in music description. We select widely employed music understanding models as baselines and evaluate their performance on MuChin. The recent music understanding models include MERT-95M, MERT-330M~\cite{li2023mert}, Jukebox-5B~\cite{jukemir}, Music2Vec~\cite{music2vec} and EnCodec~\cite{encodec}. And considering that Jukebox-5B is a pre-trained generative model, not originally designed for music understanding, we use the method in~\cite{jukemir} to encode audio with Jukebox-5B.

\subsubsection{Evaluation Metrics}

% To evaluate the performance of music understanding models, we input music audio into them, and get the encoded sequences respectively. Then for each model, we employ a classifier with an average pooling layer and 5 linear layers to extract 10 sets of music descriptive tags corresponding to the dimensions from its outputting encoded sequences.

To assess the effectiveness of music understanding models, we feed music audio into them and obtain their respective encoded sequences. Subsequently, for each model, we utilize a classifier comprising an average pooling layer and 5 linear layers to extract 10 sets of descriptive music tags corresponding to the dimensions of its output encoded sequences. %TZ

\begin{itemize}
    \item \textbf{Semantic Similarity Score.} The BGE model~\cite{bge_embedding}, as a general word vector embedding model, has demonstrated impressive performance on various tasks. We utilize the bge-large-zh-v1.5 model to calculate the semantic similarity between the generated and original tags.
\end{itemize}

For each set of test data, we can ascertain the semantic similarity between them by encoding the tags into embeddings using the BGE model and computing the outer product of these embeddings. Then we sequentially enumerate each generated tag against the original tags, calculate the Semantic Similarity Scores between them, and then obtain the average of all the values as the score of a specific model.

\subsubsection{Results}

\textbf{Table~\ref{tab:und_results}} demonstrates the semantic similarity scores of the five selected models. It can be observed that, MERT, which encodes both audio and music attributes, performs best in understanding and describing music. Thanks to its massive number of parameters and volume of training data, Jukebox also achieves commendable results. However, as its architecture does not emphasize music attributes, its performance does not reach its full potential.

Moreover, for MERT-95M and MERT-330M, despite their scores being relatively close, we still observe the inverse-scaling effect across multiple dimensions, consistent with the phenomenon mentioned in the paper of MERT~\cite{li2023mert}. Specifically, for objective music attributes such as rhythm and instrumentation, MERT-330M performs better, but for most subjective descriptive dimensions, MERT-95M shows superior performance. Therefore, we hypothesize that, in line with the descriptions in the MERT paper, as the amount of data and the number of parameters increase, MERT incorporates more music attribute information, which makes it easier for the model to extract music attributes. However it may lead to a dilution of some audio description-related information. This also indicates that the music attributes extracted by MIR cannot be directly used for music description benchmarks.

% We believe there should be a trade-off between music attributes and audio information, and that introducing a larger number of parameters and more information does not necessarily enable the model to describe music better when the model is sufficiently large.

% 同时编码音频和音乐属性的MERT在理解并描述音乐中的表现最佳。得益于大量的参数量和训练数据量，Jukebox也取得了不错的效果，但是由于架构中并没有强调音乐属性，其性能并没有达到最佳。此外，对于MERT的95M和330M两个版本，尽管二者的评分相差不大，我们还是在多个维度观察到了inverse-scaling effect，与MERT原文中提到的现象一致。具体地，对于节奏、配器等客观的音乐属性，MERT-330M表现更好，但对于大多数带有主观性的描述维度，MERT-95M的表现更好。因而我们推测，按照MERT原文中的描述，随着数据量和参数量的提升，MERT引入了更多音乐属性信息，模型更容易提取到音乐属性，但是可能会导致一些音频描述方面的信息的稀释。我们认为音乐属性和音频信息之间应该做出trade-off，并且当模型足够大时，引入更大的参数量和更多的信息不一定使模型能够更好地描述音乐。

\begin{table*}[htb!]
\centering
\resizebox{\linewidth}{!}{%
 \renewcommand\arraystretch{1.3}
% \fontsize{7}{9}\selectfont
\begin{tabular}{rrrrrrr}  
\toprule
Model&  &  Jukebox &MERT-330M&  MERT-95M &Music2Vec& EnCodec\\  \hline 
 Parameter Size&  & 5B &330M& 95M &95M& 56M\\
 Data (h)&  & 60 $\sim$ 120k  &160k & 17k  &1k & 1k\\
\hline
& Average Score-P& 0.5490(±0.1458)  &0.5586(±0.1433) & \textbf{0.5640(±0.1425)}  &0.5474(±0.1417) &  0.4583(±0.1377)\\ \cline{2-7}
 & Tempo \& Rhythm & 0.4610(±0.1016)    &\textbf{0.4650(±0.1013)} & 0.4607(±0.0958)  &0.4604(±0.1026)  & 0.4587(±0.1092)\\
 & Emo. Impact (L \& M)& 0.5312(±0.0939)  &0.5350(±0.0903) & \textbf{0.5396(±0.0857)}  &0.5311(±0.0924) &  0.4860(±0.0920)\\
 & Cult. \& Reg.& 0.5166(±0.2107)  &0.5340(±0.2139) & \textbf{0.5390(±0.2110)}
  &0.5120(±0.2094) & 0.4072(±0.1261)\\
 Professional& Vocal Components  & 0.5464(±0.1953)  &0.5550(±0.1957)&\textbf{0.5713(±0.1989)}  &0.5356(±0.1926) &0.4230(±0.1361)\\
 Description& Song Purp. &0.5810(±0.2191)  &0.5864(±0.2166) &\textbf{0.6040(±0.2230)}  &0.5664(±0.2144)&0.4630(±0.1504)\\
 & Mus. Genres&0.4600(±0.1239)  &0.4644(±0.1172) & \textbf{0.4692(±0.1158)}  &0.4610(±0.1207) &0.4297(±0.1219)\\
 &Exp. Impact (S \& A) &0.9146(±0.0541)  &0.9280(±0.0476)& \textbf{0.9310(±0.0447)} &0.9190(±0.0576) &0.7085(±0.2888)\\
 &Tgt. Aud. & 0.4521(±0.1471) &0.4656(±0.1459) &\textbf{0.4683(±0.1417)}  &0.4565(±0.1514) &0.3623(±0.0980)\\
 &Instrum. & 0.5083(±0.1647) &\textbf{ 0.5180(±0.1587)}& 0.5156(±0.1592) &0.5063(±0.1727)&0.4043(±0.1426)\\
 &Audio Eff. & 0.5195(±0.1476)  &0.5356(±0.1458) & \textbf{0.5425(±0.1483)}  &0.5244(±0.1539) &  0.4404(±0.1122)\\ \hline
& Average Score-A& 0.5894(±0.1353)  &0.5900(±0.1284) & \textbf{0.5923(±0.1284)} &0.5770(±0.1417) & 0.4602(±0.1449) \\ \cline{2-7}
 & Perc. of Tempo& \textbf{0.4600(±0.1521)}    &0.4540(±0.1475) &0.4580(±0.1456)    &0.4463(±0.1407) & 0.4065(±0.0994) \\ 
 & Emo. Impact (L)& 0.5977(±0.1780) &0.5894(±0.1798) &\textbf{0.6006(±0.1780)}
  &0.5806(±0.1827) & 0.4430(±0.1320)\\
 & Cult.\& Reg. & 0.4565(±0.1013)  &0.4539(±0.0975) & \textbf{0.4575(±0.0949)} &0.4510(±0.1023) & 0.4324(±0.0972)\\
 Amateur&  Vocal Components & \textbf{0.5195(±0.1208)} &0.5190(±0.1216)&0.5186(±0.1227)  &0.5117(±0.12 00) &0.4795(±0.0950)\\
 Description& Song Purp.& 0.5240(±0.2377) &0.5210(±0.2356)& \textbf{0.5410(±0.2422)} &0.5201(±0.2428)&0.3801(±0.1532)\\
 &Perc. of Uniq. &0.5356(±0.2076)  &0.5356(±0.2115) &\textbf{0.5547(±0.2085)}  &0.5060(±0.1942)&0.4130(±0.1191)\\
 &Exp. Impact (S) & 0.9404(±0.0328) &0.9385(±0.0315) & \textbf{0.9460(±0.0315)} &0.9297(±0.0477) &0.7144(±0.2640)\\
 &Tgt. Aud. &0.4417(±0.1041)  &0.4448(±0.1114) &\textbf{0.4530(±0.0951)}  &0.4353(±0.1220)&0.3933(±0.1075)\\
 &Instrum. &0.7144(±0.0737)  &\textbf{0.7153(±0.0537)} &0.6787(±0.0333)  &0.6807(±0.1059) &0.4219(±0.2092)\\
 &Audio Eff. & 0.7056(±0.1448)  &\textbf{ 0.7275(±0.1465)} & 0.7144(±0.1326)  &0.7110(±0.1586) & 0.5176(±0.1725) \\ \bottomrule
\end{tabular}%
}
\caption{Evaluation results of selected music understanding models on the benchmark. The metrics of description presented in the table can be referenced to the \textbf{descriptive dimensions} of P and A on the right side of \textbf{Figure~\ref{fig:MuChin}}. After encoding music by the models, we employ an MLP to output descriptive tags corresponding to these dimensions. The \textbf{pipeline} of this process can be found in \textbf{Appendix~\ref{app:und}}. The method for calculating the \textbf{semantic similarity} scores between the model's output results and the test set labels can be referenced in \textbf{Section~\ref{sec:und}}.}

\label{tab:und_results}
\vspace{-0.2cm}
\end{table*}

\section{Related Work}

\textbf{Datasets Based on MIR Algorithms.} Datasets based on MIR algorithms employ existing MIR algorithms to extract musical attributes from symbolic music or music audio. And then the attributes are either incorporated into complete descriptive texts or regarded as descriptive tags. MSD~\cite{bertin2011million} collects a million of music data, along with audio, MIDI, and tags retrieved by Echo Nest Analyze API\footnote{\url{https://developer.spotify.com/}} (MIR toolkit). POP909~\cite{wang2020pop909} presents a dataset containing audio, lead sheets, and other music attributes like keys and beats. MuseCoco~\cite{lu2023musecoco} and Mustango~\cite{melechovsky2023mustango} extract features from the original audio and then utilize ChatGPT to incorporate them as descriptions. MuLaMCap in Noise2Music~\cite{huang2023noise2music} utilizes an LLM to generate a set of music descriptive texts, and then employs MuLan~\cite{huang2022mulan}, a text-music embedding model to match these texts with the music audio in the datasets. 
    
\textbf{Datasets Based on Manual Annotation.} Some datasets based on manual annotations collect descriptions or tags from music websites, while others include data annotated by professional musicians. Hooktheory\footnote{\url{https://www.hooktheory.com/}} is a music website where users upload audio with their annotations such as melodies, chords, and beats. MTG~\cite{bogdanov2019mtg} and M\^ousai~\cite{schneider2023mo} use corresponding tags of music on music websites as descriptive tags, while ERNIE-Music~\cite{zhu2023ernie} uses comments of music as music descriptions, and establish datasets upon these. Musiclm~\cite{agostinelli2023musiclm} presents a dataset, MusicCaps, including music descriptions annotated by professional musicians.
    
\textbf{Existing Benchmarks in the Field of Music.} There are several benchmarks for specific domains in the field of music. Sheet Sage~\cite{donahue2021sheet} presents a benchmark for melody transcription. GTZAN~\cite{sturm2013gtzan} presents a test set for music genre classification. PMEmo~\cite{zhang2018pmemo} has collected music emotional annotations and simultaneous electrodermal activity signals for 794 songs, thereby providing a benchmark for music emotion recognition. MARBLE~\cite{yuan2023marble} is a comprehensive benchmark for music understanding models on 4 levels of downstream MIR tasks. However, there is a lack of comprehensive benchmarks focusing on colloquial music description.

\section{Conclusion}

% In this work, we 搭建了一个标注平台(CaiMAP)来弄好了 a music description dataset in Chinese colloquial language(CaiMD). And based on this, we established the MuChin, which offers a new perspective on the performance of language models in the field of music, requiring models not only to extract basic attributes from music and describe it from a professional standpoint but also to align with the public.

% In this work, we developed an annotation platform (CaiMAP) to create a music description dataset in Chinese colloquial language (CaiMD). Based on these, we established the MuChin benchmark, which provides a new perspective on the performance of language models in the field of music, requiring models not only to describe music from a professional standpoint but also to align with public perception.

In this study, we developed an annotation platform called CaiMAP to create a dataset of music descriptions in colloquial Chinese language, termed CaiMD. Leveraging these resources, we introduced the MuChin benchmark, which offers a novel perspective on the performance of language models in the realm of music. MuChin challenges models not only to provide professional-level descriptions of music but also to align with public perceptions. %TZ

% Although we have endeavored to make MuChin as comprehensive and holistic as possible, it only accounts for tasks related to the understanding and generation of music descriptions, and still cannot fully reflect the all-around capabilities of the models in the music domain.
Despite our efforts to make MuChin as comprehensive and inclusive as possible, it solely addresses tasks related to understanding and generating music descriptions. As such, it does not fully capture the overall capabilities of models in the field of music.  %TZ

% Although we have endeavored to make MuChin as comprehensive and holistic as possible, it only accounts for tasks related to the understanding and generation of music descriptions, and still cannot fully reflect 模型在音乐领域的全面能力.

% In the future, building upon CaiMD, we intend to investigate methods for evaluating the quality of music generated by music generation models, with the aim of establishing a benchmark for music generation.

% \clearpage

%% The file named.bst is a bibliography style file for BibTeX 0.99c
\bibliographystyle{named}
\bibliography{ijcai24}

\clearpage
\appendix

% \section{Textual Description}
% \label{app:text_dis}

% Meanwhile, we hope the dataset can have descriptions that reflect the specialized characteristics of music, as well as popular, mainstream descriptions. Based on this consideration, we plan to categorize the annotators into professional and amateur groups according to their levels of musical expertise. The professional group will be responsible for providing more specialized descriptions of the data, such as clearly labeling the musical genres, instrumentation, rhythmic characteristics, and musical forms. The amateur group will be responsible for providing more popular, mainstream descriptions, including colloquial language and internet slang. To cater to the different focuses of the professional and amateur groups, we established different libraries for each group: the amateur group uses a library consisting of popular tags collected from the Internet, while the professional group uses a library constructed based on the descriptions in MusicCaps~\cite{agostinelli2023musiclm}.

\section{Tasks with Automatic Annotation}
\label{app:auto_anno}

The 
performance of current algorithms designed for annotating %TZ
textual description, lyrics, and musical section annotation is 
not satisfactory due to their reliance on subjective human evaluation.  %TZ
%However, other types of information, such as phonetic alignment, vocal separation, and audio-to-MIDI transcription, bear no significant correlation to human perception. Moreover, annotating these aspects is challenging for humans, necessitating substantial effort and time. Currently, there is an abundance of mature algorithms capable of accomplishing these tasks, which will be discussed in \textbf{Appendix~\ref{app:data_pro}}. Consequently, we employ the data preprocessing algorithms for the automatic annotation of this content, foregoing manual annotation or intervention, and integrate it directly into our dataset.
Except these, other kinds of data, such as phonetic alignment, vocal separation, and audio-to-MIDI conversion, do not significantly align with human perception. Annotating these elements manually is particularly challenging, requiring extensive effort and time. However, there are now numerous advanced algorithms that efficiently handle these tasks, as detailed in the \textbf{Appendix~\ref{app:data_pro}}. As a result, we utilize data preprocessing algorithms for the automatic annotation of this content, eliminating the need for manual annotation or intervention, and seamlessly integrate this processed content into our dataset.  %TZ

% 文本描述、歌词押韵、乐段结构使用既定算法效果不好，依赖人类的主观感受判别，而除此之外的信息，如字-音符对齐、音频分离、音频-符号转写midi，与人类感受并无明显关系，并且这部分内容的标注人类难以胜任，或费时费力。目前这部分处理已经有大量成熟的算法可以完成，在Section 3.2.1会对这些算法有详细的叙述。因此这部分内容使用成熟的数据预处理算法进行自动化标注，不使用人工进行标注或干预，直接纳入数据库中。

% Please add the following required packages to your document preamble:
% \usepackage{multirow}
% \usepackage{graphicx}
\begin{table}[htbp]
\centering
\fontsize{9}{10}\selectfont
\renewcommand\arraystretch{1.1}
\begin{tabular}{cc}
\hline
\textbf{Classification} & \textbf{Task}                  \\ \hline
\multirow{4}{*}{A}      & Musical Section Annotation     \\
                        & Lyric Correction               \\
                        & Lyric Screening                \\
                        & Rhyme Annotation               \\ \hline
\multirow{2}{*}{B}      & Professional Music Description \\
                        & Amateur Music Description      \\ \hline
\end{tabular}%
\caption{Classification of Annotation Tasks.}
\label{tab:class_task}
\vspace{-0.3cm}
\end{table}

\section{Data Preprocessing}
\label{app:data_pro}

\begin{itemize}
    % 跟正文中2.1.1的Textual Description第二段重复了
    % \item \textbf{Lexicon of Descriptive Terms} To enhance the efficiency, accuracy, and consistency of annotations, and to align with the public while also facilitating users, we build a lexicon of descriptive terms. This lexicon is divided into a popular term lexicon and a musical terminology lexicon. The former consists of popular music descriptive terms collected from the internet and sorted by frequency, while the latter contains keywords extracted from the descriptive texts of the open-source text-music dataset MusicCaps~\cite{agostinelli2023musiclm}.
    
    \item \textbf{Music Genre Clustering} %To prevent subjective bias and a lack of diverse descriptions for certain music genres, it is essential to allocate a wide range of music genres to each annotator, ensuring the diversity of the annotations. To achieve this, we employ MERT~\cite{li2023mert}, a pre-trained music audio encoder, to encode the audio data. Subsequently, we perform clustering on the encoded data, resulting in 1000 distinct audio clusters. We then extract music data from these clusters evenly, ensuring that annotators receive a balanced selection of music for labeling. By implementing this method, we ensure that each music cluster receives a sufficient number of descriptions from different annotators, thereby enhancing the diversity of the annotated data.
    To mitigate subjective bias and ensure diverse descriptions across various music genres, it's crucial to distribute a broad spectrum of music genres among annotators, thereby enriching the annotation's diversity. To facilitate this, we utilize MERT~\cite{li2023mert}, a pre-trained music audio encoder, to process the audio data. Following this, we cluster the encoded data, resulting in 1000 unique audio clusters. From these clusters, we evenly distribute music data, guaranteeing that annotators are presented with a balanced mix of music for labeling. This approach ensures that each music cluster is described by a range of annotators, significantly enhancing the diversity and richness of the annotated data. %TZ
    
    \item \textbf{Vocal \& Track Separation} To make the dataset suitable for tasks such as accompaniment generation, melody generation, and vocal synthesis, we apply Demucs ~\cite{rouard2022hybrid,defossez2021hybrid} to perform vocal separation, separating the vocals from the musical accompaniment in the audio files. Furthermore, considering the requirements of a wider range of music-related tasks, we also separate individual instrument tracks, such as drums and bass.
    
    \item \textbf{Phonemic Level Alignment in Audio-Lyrics} %For audio-lyrics pairs, we need to perform phonemic level alignment to make them suitable for tasks like vocal synthesis. We utilize the Montreal Forced Aligner (MFA)~\cite{mfa} to align the audio-lyrics pairs, achieving an accuracy of 67\%. While MFA achieves a high accuracy of 95\% for aligning single-pitch phonemes with single characters, it tends to incorrectly mark the offsets for melismatic phonemes, which involve singing multiple pitches within a single syllable or note, resulting in lower overall accuracy. To address this, we optimize the MFA algorithm by prioritizing the identification and alignment of melismatic phonemes. Additionally, we recognize and mark long pauses and breaths that occur during singing. With these improvements, we achieve a final alignment accuracy of 97\%.
    To prepare audio-lyrics pairs for applications such as vocal synthesis, it's necessary to align them at the phonemic level. We employ the Montreal Forced Aligner (MFA)~\cite{mfa} for this task, initially achieving a 67\% accuracy rate. While the MFA demonstrates a commendable 95\% accuracy for aligning monophonic phonemes to single characters, its performance drops due to inaccuracies in marking the offsets for melismatic phonemes. These phonemes are characterized by multiple pitches sung within a single syllable or note, complicating the alignment process and diminishing the overall accuracy. To address this, we optimized the MFA algorithm with a focus on accurately identifying and aligning melismatic phonemes. Furthermore, we implemented features to recognize and annotate significant pauses and breaths during singing. These enhancements significantly improve our final alignment accuracy to 97\%.  %TZ
    
    \item \textbf{Automatic Pre-annotation} %In order to enhance the efficiency of subsequent manual annotation, we employ predetermined programs for automatic pre-annotation on certain lyric annotation tasks. For lyric rhyme annotation, a identification program is utilized to pre-annotate the rhyme scheme of each line, while in lyric theme annotation, the fine-tuned Qwen is utilized to extract the main theme of lyrics for each music piece in advance. During the formal annotation phase, all pre-annotation results serve as references for manual annotation. Annotators can verify the accuracy of the pre-annotation and make modifications based on it, or use the pre-annotation results as references for their own annotation tasks.
    To improve the efficiency of future manual annotations, we implemented specific software for automatic pre-annotation of certain tasks related to lyric annotation. For annotating rhyme schemes in lyrics, we use a specialized program that pre-annotates the rhyme scheme for each line. For theme annotation in lyrics, we employ a fine-tuned version of Qwen to preliminarily identify the main theme of the lyrics for each piece of music. During the formal annotation phase, these pre-annotations serve as a basis for manual review. Annotators can assess the accuracy of these automatic annotations and adjust them as necessary, or use them as a guideline for their own annotation efforts.  %TZ
    
    \item \textbf{Lead Sheet Transcription} To facilitate symbolic music-related tasks using MIDI, we transcribe the audio in the MuChin into lead sheets. % a simple form of MIDI notation, using Sheet Sage~\cite{donahue2021sheet}, which is based on the encoding of Jukebox~\cite{dhariwal2020jukebox}. This enables its application to tasks related to symbolic music.
    These sheets, which are a simplified form of MIDI notation, are created using Sheet Sage~\cite{donahue2021sheet}, software that utilizes the encoding model of Jukebox~\cite{dhariwal2020jukebox}. This conversion facilitates the application of MuChin to a wide range of tasks associated with symbolic music. %TZ
\end{itemize}

% 聚类 为了避免每个标注员分配到的音乐风格数量不平衡，导致主观偏差，即某些音乐风格无法收集到多样化的描述，我们需要尽量为每个标注员分配多样的不同风格的音乐，从而保证数据标注的多样性。因此，我们预先使用MERT[2]将数据编码，对音乐数据编码进行聚类，得到1000类音乐，然后尽量均匀地从这些类中抽取音乐数据提供给标注员进行标注，以此使每类音乐都能得到足够多不同标注员的描述，从而提升标注数据的多样性。

% 人声&音轨分离 为了数据集能够用于伴奏生成、旋律生成、歌声合成等任务，我们使用demucs算法（加引用）对完整音乐音频进行了人声分离，得到纯人声和纯伴奏音频。更进一步地，考虑到更多的音乐任务，一些乐器的音轨可以被进一步分离，例如鼓、贝斯。

% 音频-歌词对齐 对于音频-歌词pair，我们需要进行音素级别的对齐，使其能够用于歌声合成等任务中。MFA对单音-单字的对齐能够达到95%的准确率，然而对于转音字，会错误地标记该字的offset，导致使用MFA算法对音频歌词对进行对齐整体准确率不高（67%的准确率）。因此，我们在MFA基础上进行优化，优先识别并标记转音字。同时，人在歌唱时常会有长时间的停顿、喘息等，我们将这些部分也进行了识别与标记。进行上述处理后，最终得到97%的对齐准确率。

% 自动预标注 为了提升后续人工标注的效率，我们使用预设程序及微调后的Qwen[3]对一些歌词标注任务进行了自动预标注。在歌词韵律标注中，我们使用程序预先标注了每句话的韵脚；在歌词主旨标注中，我们使用Qwen预先提取了每个歌词的主旨。在正式标注阶段，所有预标注任务都仅作为人工标注的参考，数据标注员可以检查预标注的正确性，并在此基础上进行修改，或参考预标注结果进行自己的标注任务。

% Lead Sheet Transcription 为了适配使用MIDI的符号音乐相关任务，我们使用基于Jukebox [3]编码的Sheet Sage [4]将数据集中的音频转写成Lead Sheet，一种简单的MIDI乐谱，使之可以应用于简单的符号音乐相关任务。

\begin{table}[htbp]
\centering
\fontsize{9}{10}\selectfont
\begin{tabular}{@{}rr@{}}
\toprule
\textbf{Classification} & \textbf{Accuracy(\%)} \\ \midrule
I                       & {[}90, 100{]}         \\
II                      & {[}70, 90)            \\
III                     & {[}60, 70)            \\
IV                      & {[}0, 60)             \\ \bottomrule
\end{tabular}%
\caption{Classification of Annotators of Type A Tasks Based on Accuracy}
\label{tab:class_a}
\vspace{-0.3cm}
\end{table}

\begin{table}[htbp]
\centering
\fontsize{9}{10}\selectfont
\begin{tabular}{@{}rr@{}}
\toprule
\textbf{Classification} & \textbf{Score} \\ \midrule
I                       & {[}90, 100{]}  \\
II                      & {[}70, 90)     \\
III                     & {[}60, 70)     \\
IV                      & {[}0, 60)      \\ \bottomrule
\end{tabular}%
\caption{Classification of Annotators of Type B Tasks Based on Score }
\label{tab:class_b}
\vspace{-0.3cm}
\end{table}

% 这里放两个表格, 每个12行3列, 主要是专业和业余的评分细则

% Please add the following required packages to your document preamble:
% \usepackage{booktabs}
%%%%%% 专业
\begin{table*}[htb!]
\centering
\resizebox{\textwidth}{!}{
\begin{tabular}{@{}rrc@{}}
\toprule
\textbf{Dimension}  & \textbf{Score} & \textbf{Standard}                                               \\ \midrule
Expressive Impact (S. \& A.) & 13 & 4 for Number of Labels; 4 for Label Relevance; 5 for Innovation                                            \\
Emotional Impact    & 13             & 4 for Number of Labels; 4 for Label Relevance; 5 for Innovation \\
Textual Description & 8              & 3 for Description Relevance; 5 for Word Counts and Innovation   \\
Musical Genres      & 8              & 8 for Level of Detail                                           \\
Tempo and Rhythm    & 5              & 5 for Label Relevance                                           \\
Instrumentation                              & 12 & 5 for Number of Labels; 3 for Label Relevance; 2 for Description Relevance; 2 for Description Thoroughness \\
Song Purpose        & 6              & 3 for Label Relevance; 3 for Innovation                         \\
Culture and Region  & 6              & 3 for Label Relevance; 3 for Innovation                         \\
Target Audience     & 6              & 3 for Label Relevance; 3 for Innovation                         \\
Vocal Components                             & 12 & 5 for Number of Labels; 3 for Label Relevance; 2 for Description Relevance; 2 for Description Thoroughness \\
Audio Effects       & 5              & 5 for Label Relevance                                           \\
Lyric Themes        & 6              & 3 for Label Relevance; 3 for Innovation                         \\
Total               & 100            & -                                                               \\ \bottomrule
\end{tabular}
}
\caption{Scoring Guidelines of Professional Music Description}
\label{tab:standard-p}
\end{table*}

%%%%%% 业余
% Please add the following required packages to your document preamble:
\begin{table*}[htb!]
\centering
\resizebox{\textwidth}{!}{%

\begin{tabular}{@{}rrrr@{}}
\toprule
\textbf{Dimension}         & \textbf{Score} & \multicolumn{2}{c}{\textbf{Standard}}                                               \\ \midrule
Perception of Uniqueness   & 8              & \multicolumn{2}{c}{4 for Label Relevance; 4 for Innovation}                         \\
Perception of Tempo        & 5              & \multicolumn{2}{c}{3 for Label Relevance; 2 for Innovation}                         \\
Expressive Impact (S.) & 13             & \multicolumn{2}{c}{4 for Number of Labels; 4 for Label Relevance; 5 for Innovation} \\
Emotional Impact (L.)  & 13             & \multicolumn{2}{c}{4 for Number of Labels; 4 for Label Relevance; 5 for Innovation} \\
Textual Description        & 8              & \multicolumn{2}{c}{3 for Description Relevance; 5 for Word Counts and Innovation}   \\
Instrumentation  & 12 & \multicolumn{2}{c}{5 for Number of Labels; 3 for Label Relevance; 2 for Description Relevance; 2 for Description Thoroughness} \\
Song Purpose               & 6              & \multicolumn{2}{c}{3 for Label Relevance; 3 for Innovation}                         \\
Culture and Region         & 6              & \multicolumn{2}{c}{3 for Label Relevance; 3 for Innovation}                         \\
Target Audience            & 6              & \multicolumn{2}{c}{3 for Label Relevance; 3 for Innovation}                         \\
Vocal Components & 12 & \multicolumn{2}{c}{5 for Number of Labels; 3 for Label Relevance; 2 for Description Relevance; 2 for Description Thoroughness} \\
Audio Effects              & 5              & \multicolumn{2}{c}{5 for Label Relevance}                                           \\
Lyric Themes               & 6              & \multicolumn{2}{c}{3 for Label Relevance; 3 for Innovation}                         \\
Total                      & 100            & \multicolumn{2}{c}{-}                                                               \\ \bottomrule
\end{tabular}
}
\caption{Scoring Guidelines of Amateur Music Description}
\label{tab:standard-a}
\end{table*}

\section{Quality Assurance Mechanisms}
\label{app:control}

In this section, we will provide a detailed introduction to the quality assurance mechanism, including the classification of tasks, scoring guidelines and the classification of individuals.

\subsection{Classification of Annotation Tasks}
\label{app:task_class}

%Based on whether the annotation tasks can be objectively assessed, we categorize them into \textbf{Type A (yes)} and \textbf{Type B (no)}. 
We classify annotation tasks into two categories based on their potential for objective evaluation: \textbf{Type A}, which can be objectively assessed, and \textbf{Type B}, which are subject to subjective assessment. %TZ
This section exemplifies the classification of each annotation task. To maximize the accuracy and comprehensiveness of each song's annotations, we allocate two annotators to Type A tasks and one annotator to Type B tasks for each song. These tasks are carried out separately, not simultaneously. Additionally, apart from annotators, several quality assurance inspectors are needed to evaluate the annotators' outputs. According to the division into Type A and B, we consolidate Type A tasks into one phase, denoted as the \textbf{Structure Annotation Phase}, and Type B tasks into the subsequent phase, denoted as the \textbf{Music Description Annotation Phase}. Data must sequentially pass through these two phases before inclusion in the dataset. That is, data must undergo structure annotation and pass quality assurance before proceeding to the music description annotation phase, after which, data that passes quality assurance following music description annotation can be added to the dataset. For Type A tasks, if both annotators provide identical annotations, we consider the annotation accurate. However, when there is a discrepancy, quality assurance inspectors must deliver their judgment to determine which result is correct, or if both are incorrect, provide their own accurate annotation. For Type B tasks, quality assurance inspectors are required to assign a score ranging from 0 to 100 to the annotation results, with the scoring guidelines detailed in Table~\ref{tab:standard-p} and~\ref{tab:standard-a}.

% 这里放1个表格

% 首先，我们根据标注内容是否能够进行客观评价将标注任务分为A类（是）和B类（否）。Table 3-2给出了各个标注任务A, B分类。为尽可能提升每首歌标注的准确性、全面性，每首歌的A类任务安排两名标注员进行标注，每首歌的B类任务安排一名标注员进行标注（分开进行，并不是同时）。此外，除标注人员外，还需要若干质检员，负责评价标注人员的结果。我们根据A, B类任务划分，将A类任务统一在一个阶段，称为结构标注阶段；B类任务统一在下一个阶段，称为描述标注阶段。一个数据依次通过两个阶段的标注才能够被加入数据集。即，需要先执行A类标注任务，并质检通过之后才会进行B类标注，经B类标注后的数据在质检通过后加入数据集。对于A类任务，如果两个标注员是相同的标注，我们认为该标注是准确的；而对存在分歧的结果，质检员需要给出自己的判断，判定哪个结果是正确的，或者两个结果均不正确并给出自己的准确标注。对于B类任务，质检员需要对标注结果给出0~100分的评分，评分细则如Table 3-3所示。

\subsection{Classification of Individuals}
\label{app:ind_class}

%To ensure that annotators perform their tasks diligently, we design a screening mechanism for the annotators. During the structure annotation phase, the accuracy of annotations for Type A tasks is evaluated using the quality assurance mechanism mentioned previously. During the music description annotation phase, since Type B tasks are subjective descriptions and difficult to objectively judge as correct or incorrect, we randomly select 20\% of the data annotated by each annotator for quality assurance scoring. Additionally, behaviors detected by the backend, such as the frequency of the interactions with the progress bar or skipping through tasks, are also evaluated. Annotators deemed to be perfunctory will receive warnings. In both phases, annotators are classified into four types based on their weekly accuracy rate or average score, as shown in Table~\ref{tab:class_a} and~\ref{tab:class_b}. Type IV annotators, as well as those who have accumulated two or more warnings, will no longer participate in the following annotation tasks, and their data for the current week will be invalidated. Type I annotators will receive rewards, while Type III annotators will face certain penalties.
To ensure diligent performance from annotators, we have implemented a screening mechanism. During the structural annotation phase, the precision of Type A task annotations is assessed through the previously mentioned quality assurance system. In the music description annotation phase, given that Type B tasks involve subjective descriptions challenging to assess objectively, we randomly review 20\% of the annotations from each annotator for quality control. Moreover, we evaluate behaviors indicated by backend analytics, such as interaction frequency with the progress bar and task skipping. Annotators showing superficial engagement will be warned. In both phases, annotators are categorized into four groups based on their weekly accuracy rates or average scores, as detailed in Table~\ref{tab:class_a} and~\ref{tab:class_b}. Type IV annotators, and those receiving two or more warnings, will be excluded from future tasks, and their data for the current week will be disregarded. Type I annotators will be rewarded, while Type III annotators may incur penalties. %TZ

% 为了保证标注人员认真进行标注，我们设计了一套筛选机制，用于标注人员的筛选。在词 结构标注阶段，标注员对A类任务的标注会经由前文所述的质检员机制判断正确率，而在描述标注阶段，B类任务由于主要是主观描述，难以客观判断正确与否，因此我们会随机抽取每个标注员标注数据的20%交由质检员评分。另外，后台监测到的标注员行为，例如拖动进度条的频率、跳过频率等也会被评判，行为被管理员认定为敷衍标注的标注员，会受到警告。在两个阶段，分别根据每周正确率和平均评分将标注员分为4类，如Table 3-4 和 3-5 所示。IV类及累计受到两次警告的标注员不再参与下周的标注任务，且本周的标注数据作废。I类标注员会得到一些奖励，III类标注员则会受到一定惩罚。

% 这里放两个表格, 分级处理

\subsection{Other Quality Assurance Measures}
\label{}

Annotators are responsible for screening the data (Type A \& B). For songs that contain languages other than Chinese, have poor audio quality, or involve pornography or violence, therefore unsuitable for inclusion in the dataset, annotators can mark these for exclusion and skip their annotation.

%When annotating musical sections (Type A), annotators are required to listen to a piece of music repeatedly. Therefore, we judge the seriousness of their annotation efforts based on the duration of time they spend on the annotation page, the frequency of their interactions with the progress bar, and how often they click the play/pause button.
When annotating musical sections of Type A, annotators must repeatedly listen to a music piece. Consequently, the dedication to their annotation tasks is assessed by the amount of time they spend on the annotation page, their frequency of interactions with the progress bar, and the frequency of their play/pause button clicks. %TZ

In the textual description annotation (Type B), to ensure that annotators listen to each song attentively and provide thoughtful music descriptions, we stipulate that annotators must listen to the entire song in one sitting before adjusting the progress bar and playback speed. They must compose a textual description of no fewer than 50 words, and are prohibited from writing the description within the first 30 seconds of the song's playback, as well as from copying and pasting any content.

% 标注员需要对数据进行筛选。对于音质和听感过差、包含色情暴力等因素不宜进入数据集的歌曲，标注员可以标记并跳过该数据的标注。
% 标注乐段时需要标注人员反复听一段音乐，所以我们根据标注人员页面停留时间，拖动进度条及点击播放/暂停键的频率，判断其标注的认真程度。
% 在文本描述标注中，为了保证标注人员认真地听完每首歌曲并认真给出音乐描述，我们要求标注人员必须先一次性完整地听完整首音乐才能对进度条和播放速度进行调整，填写不少于50字的文本描述，且在音乐开始播放的30s内不能填写描述，也不能复制粘贴。

\begin{figure}[htb!]
    \centering
    \includegraphics[width=0.3\textwidth]{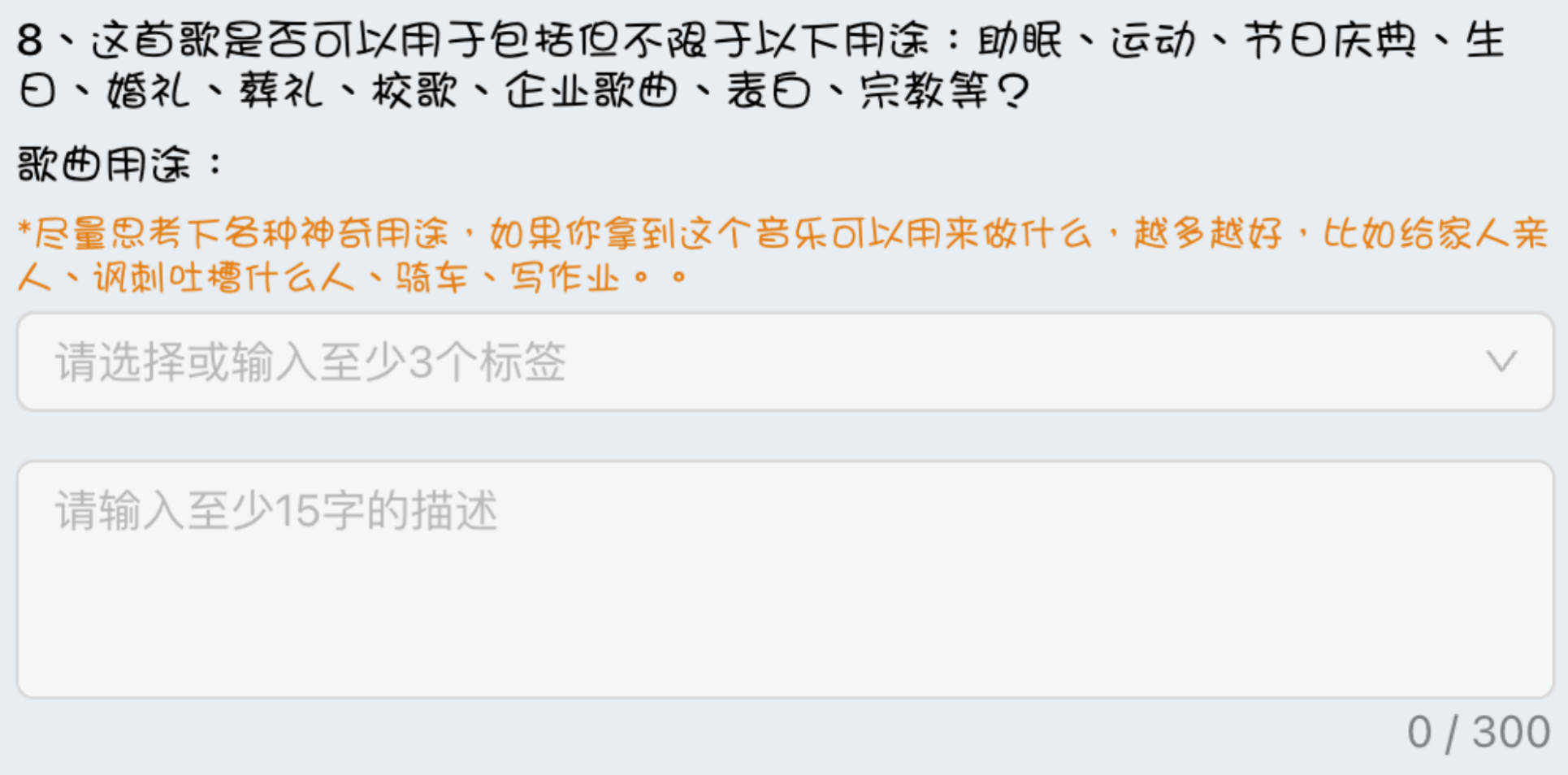}
    \caption{Supplementary actual screenshots from the main text. A screenshot of the 'Song Purpose' section during the Description Annotation Phase.}
    \label{fig:caimap-5}
\end{figure}
\vspace{-0.4cm}

\begin{figure}[htb!]
    \centering
    \includegraphics[width=0.3\textwidth]{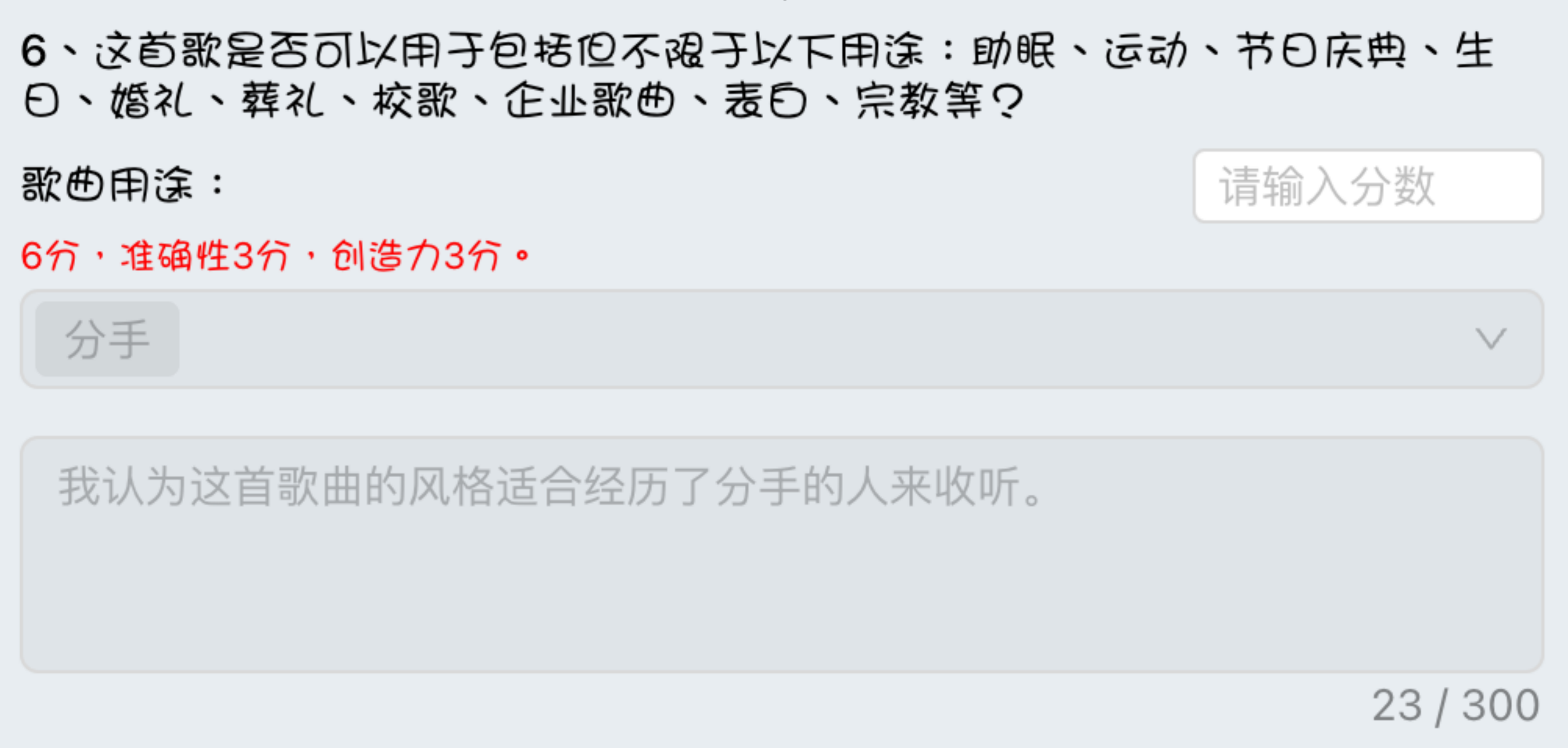}
    \caption{Supplementary actual screenshots from the main text. A screenshot of the 'Song Purpose' section during the Description Quality Assurance Phase.}
    \label{fig:caimap-1}
\end{figure}
\vspace{-0.4cm}

\begin{figure}[htb!]
    \centering
    \includegraphics[width=0.3\textwidth]{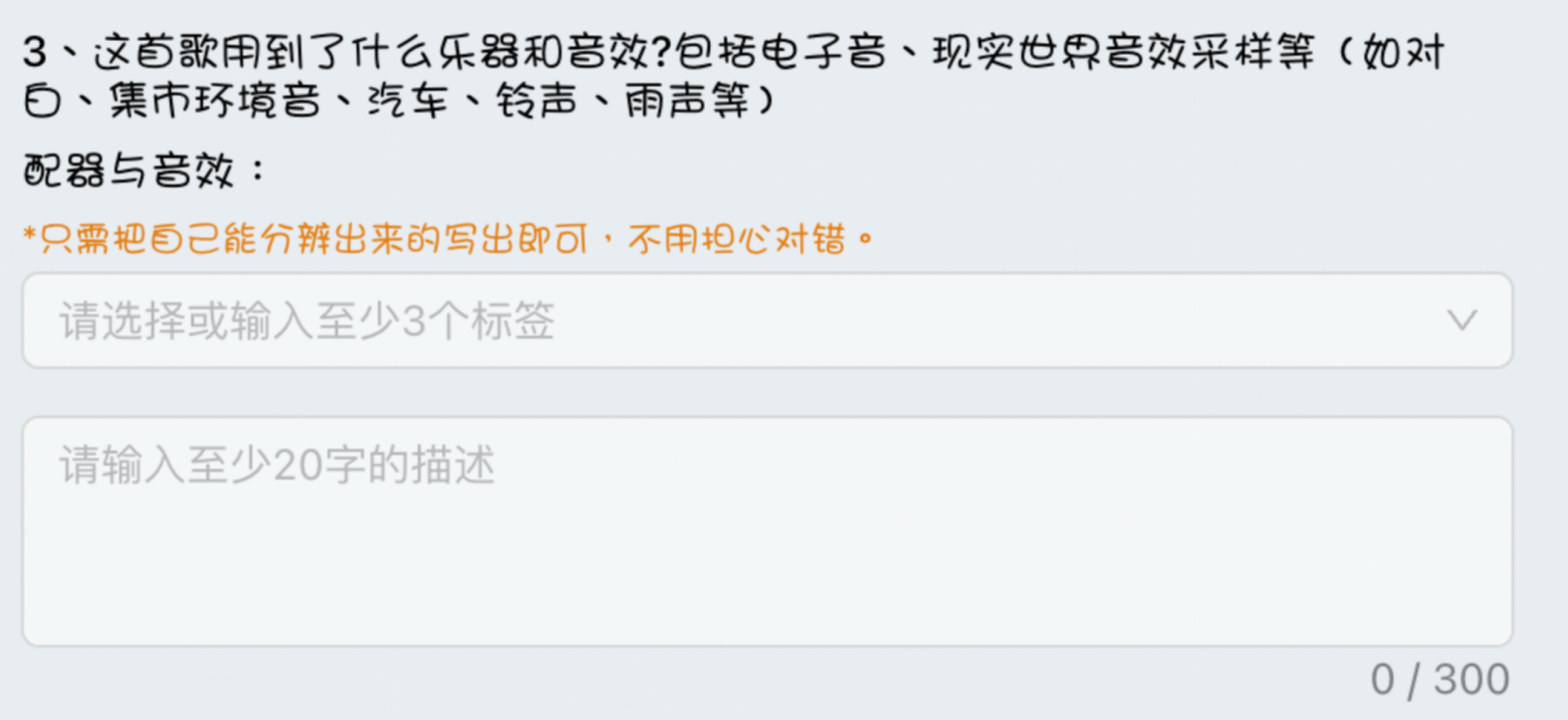}
    \caption{Supplementary actual screenshots from the main text. A screenshot of the 'Instrumentation' section during the Description Annotation Phase.}
    \label{fig:caimap-3}
\end{figure}
\vspace{-0 cm}

\begin{figure}[htb!]
    \centering
    \includegraphics[width=0.3\textwidth]{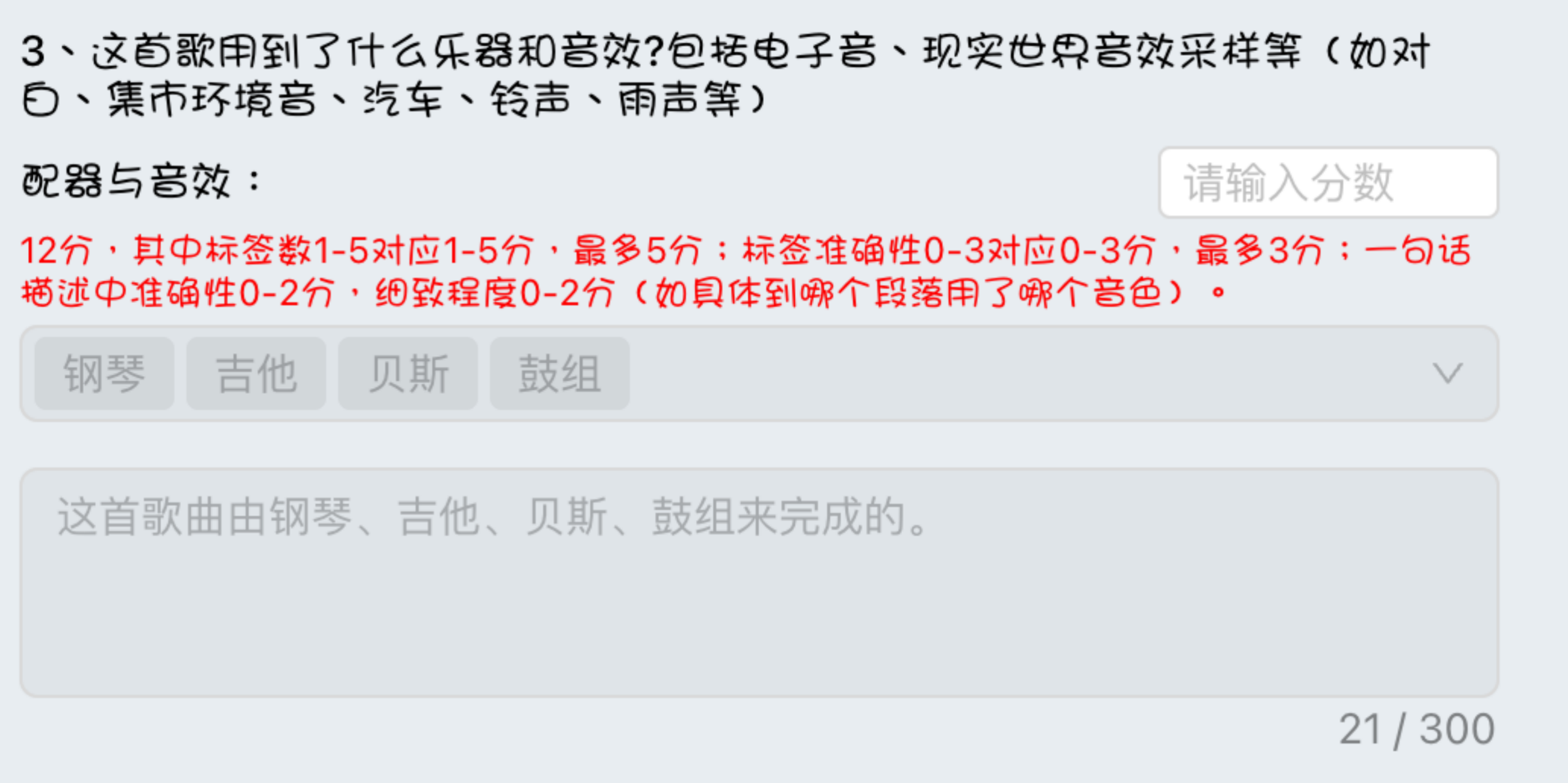}
    \caption{Supplementary actual screenshots from the main text. A screenshot of the 'Instrumentation' section during the Description Quality Assurance Phase.}
    \label{fig:caimap-2}
\end{figure}
\vspace{-0.4cm}

\begin{figure}[htb!]
    \centering
    \includegraphics[width=0.3\textwidth]{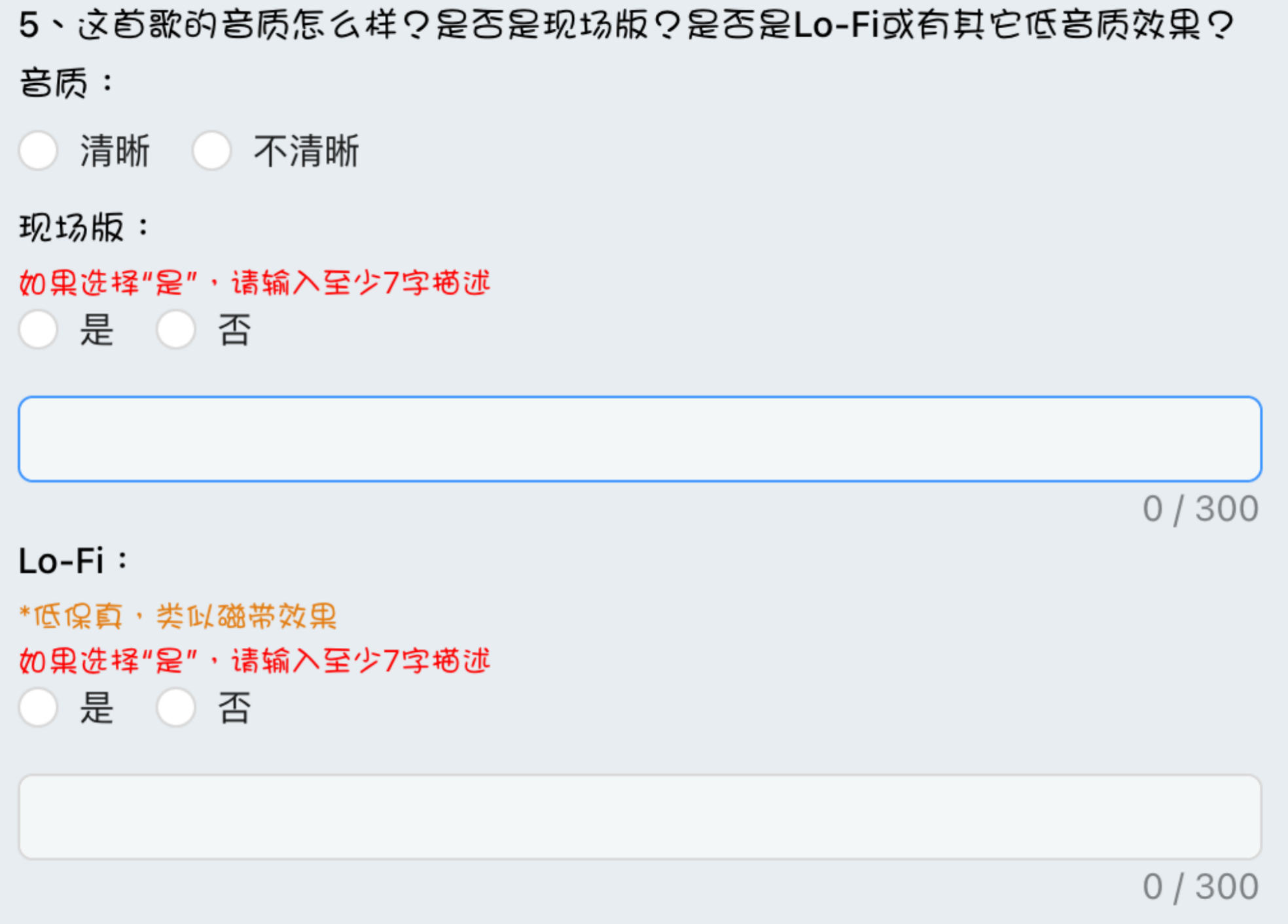}
    \caption{Supplementary actual screenshots from the main text. A screenshot of the 'Audio Effects' section during the Description Annotation Phase.}
    \label{fig:caimap-4}
\end{figure}
\vspace{-0.4cm}

\section{CaiMAP: Caichong Multitask Music Annotation Platform}
\label{app:caimap}

In Appendix~\ref{app:control}, %we have introduced an array of annotation tasks and a sophisticated quality assurance mechanism. To actualize these intricate designs, we developed the Caichong Multitask Music Annotation Platform (CaiMAP), to integrate the series of tasks and mechanisms. The platform will be introduced in brief in this section.
we have launched a comprehensive suite of annotation tasks alongside an advanced quality assurance system. To bring these complex designs to life, we developed the Caichong Multitask Music Annotation Platform (CaiMAP), which harmonizes this series of tasks and systems. This section will provide a brief overview of the platform.

\begin{itemize}
    \item \textbf{Account and Login.} %The platform employs a username and password system for access, with accounts distributed to users based on the specific nature of their tasks. Each account is tethered to a distinct task, necessitating users to utilize the allocated account to accept, execute, and submit the pertinent task.
    The platform utilizes an access control system, assigning specific roles to each user account. Users can log into their accounts, review and complete assigned tasks, and submit their results.  %TZ
    
    \item \textbf{Annotation Interface.} %Annotators, upon logging in and selecting a specific music piece, are directed to its dedicated annotation interface. This interface is equipped with a media player and a text box tailored to the task. Users can adjust the player's progress bar and the speed of playback. And the music description annotation page features an integrated lexicon and search utility, allowing users to choose appropriate descriptive terms from the lexicon or to search for the desired terms.
    Upon logging in and selecting a specific piece of music, annotators are directed to a dedicated annotation interface designed for the task. This interface includes a media player and a specialized text box. Users have the ability to control the progress bar and playback speed of the media player. Furthermore, the music description annotation interface incorporates a comprehensive lexicon and search tool, enabling users to select suitable descriptive terms directly from the lexicon or to search for specific terms as needed.  %TZ
    
    \item \textbf{Quality Assurance Interface.} % Quality assurance inspectors, upon logging in and selecting a specific music piece, are directed to its dedicated quality assurance interface. This interface adapts its presentation to the specific quality assurance task. For Type A tasks, inspectors are tasked with evaluating the annotations of two users concurrently. In such cases, the interface displays the annotations side by side, delineating the differences for review. Inspectors may determine one of the annotations as correct, make modifications to either, or opt to re-annotate correctly themselves by selecting the re-annotate option. For Type B tasks, the quality assurance interface exhibits a single complete annotation for the inspector’s assurance and scoring. The inspectors simply review the annotations and submit their scores.
    Upon logging in and selecting a specific piece of music, quality assurance inspectors are taken to the quality assurance interface. For Type A tasks, inspectors are responsible for simultaneously evaluating the annotations provided by two users. The interface presents these annotations side-by-side, highlighting the differences for easy comparison. Inspectors can then decide which annotation is correct, make adjustments to either, or choose to re-annotate the piece. For Type B tasks, the interface displays a single, complete annotation for the inspector to verify and score. Inspectors simply review the annotation and submit their scores. %TZ
    
    \item \textbf{Administrator Interface.} Administrators have the access to view the submissions of any designated user, including annotators and quality assurance inspectors. Both the annotation and quality assurance interfaces incorporate a feedback button for reporting platform issues, enabling annotators and quality assurance inspectors to communicate with administrators for resolution.
\end{itemize}

We have provided screenshots of several platform pages as examples, as shown in Figures~\ref{fig:caimap-5}--\ref{fig:caimap-4}.

% - 账号与登录 平台采用账号密码登录，每个用户会被依据任务的不同分发账号。每个账号都会与一个特定的任务绑定，用户需要使用被分发的特定账号领取特定任务，完成并提交。
% - 标注页面 标注员账号登录后点击特定歌曲后会进入该歌曲的标注页面。平台的标注页面设有一个播放器以及指定任务对应的标注文本框。播放器进度条和播放速度可以由用户调整。在结构标注页面，内置了自动预标注的主旨结果，供用户参考。在文本描述标注页面则内置了描述词库和搜索功能，用户可以在词库中挑选符合需求的描述词，或使用关键词搜索功能检索词库查找所需的词。
% - 质检页面 质检员账号在登录后可以点击歌曲进入质检页面。质检页面会根据质检任务的不同采用不同的呈现方式。对于A类任务，质检员需要同时检查两个用户的标注。对于此类任务，质检页面会并列呈现两个用户的标注，并标记两个结果的不同之处，供质检员检查。质检员可以选择其中的一个作为正确标注提交，也可以对二者进行修改，也可以点击重新标注按钮，自行标注一份正确的标注。对于B类任务，质检页面会完整地呈现一份标注，供质检员检查并打分，质检员只需浏览该标注结果并提交评分。
% - 管理员页面 管理员账号拥有查看任意指定用户提交结果的权限，包括标注员和质检员。同时标注和质检页面中均有问题反馈按钮，标注员和质检员可以点击按钮反馈平台存在的问题，以联系管理员解决。

\section{Individual Grouping and Training}
\label{app:individual}

\subsection{Grouping}

% Given that each data require double annotations during the structure annotation phase, composed of Type A tasks, and only a single annotation during the music description annotation phase, composed of Type B tasks, fewer participants are involved in the music description annotation phase. The musical section annotation task of the lyric annotation phase necessitates a basic understanding of music theory; therefore, only the 104 professionals participate. From these, 11 individuals with a high level of expertise and a conscientious attitude are selected as quality assurance inspectors through resume screening and subsequent assessments, while the remaining 93 serve as annotators.
During the structure annotation phase, which consists of Type A tasks, each piece of data requires two annotations. In contrast, the music description annotation phase, made up of Type B tasks, necessitates only one annotation. As a result, the latter phase involves fewer participants. The task of annotating the musical sections in the lyric annotation phase demands a basic knowledge of music theory. Consequently, only 104 professionals are engaged in this task. Out of these, 11 individuals, distinguished by their high level of expertise and conscientious approach, are chosen as quality assurance inspectors. This selection process involves screening their resumes and conducting further assessments. The remaining 93 individuals function as annotators. %TZ

During the music description annotation phase, the 109 amateurs form the amateur group, and the 93 professionals from the previous phase form the professional group. Additionally, the 11 inspectors from the previous phase continue to serve as inspectors in this phase. Beyond the roles of annotators and quality assurance inspectors, we also select a member from our research team who is adept at using the platform, with a high level of expertise, and with strong communication skills to act as the platform administrator. % Specific task assignments for each group are as follows.

% 由于由A类任务组成的结构标注阶段中，每个数据需要被标注两次，而由B类任务组成的描述标注阶段中，每个数据仅需被标注一次，因此参与描述标注阶段的人员较结构标注阶段更少。结构标注阶段中的乐段标注任务要求参与者有一定音乐理论基础，因此只由104名专业人士参与。其中按照简历和后续考核筛选出11名专业水平高、态度认真的人员作为质检员，其余93名人员作为标注员。而在描述标注阶段，109名业余爱好者组成通俗描述组，上一个阶段的93名标注员组成专业描述组, 上一阶段的11名质检员继续担任本阶段的质检员。除标注员和质检员外，我们也从我们的团队中挑选了3位精通平台使用、专业水平高且擅长沟通的人员作为系统管理员。各组人员的具体任务分配如下。

\subsection{Training}

% Next, we provide training for the annotators and quality assurance inspectors on their respective tasks. On the one hand, each annotator logs in to CaiMAP and pre-annotate a small clustered dataset of approximately 20 entries, including the tasks of Type A and B. During this period, the annotators should familiarize themselves with the platform's functionalities and understand how to properly execute the annotation tasks. We also provide targeted training for the annotators on common errors, such as removing irrelevant information from the lyric texts and clearly marking each interjection.
Next, we offer training for both the annotators and quality assurance inspectors, focusing on their specific roles. Initially, each annotator accesses CaiMAP to pre-annotate a compact dataset of around 20 entries, which encompasses tasks of both Type A and B. This phase allows annotators to acquaint themselves with the platform's features and learn the correct procedures for completing annotation tasks. Additionally, we provide specialized training to address common mistakes, such as the elimination of extraneous information from lyric texts and the accurate identification of each interjection. %TZ

% On the other hand, training for the inspectors is somewhat more complex: not only do they need to become proficient in using the platform, but they must also establish a set of unified evaluation criteria. We collect data annotated by the annotators during the pre-annotation phase and distribute this identical dataset among all the inspectors. For the lyric annotation phase, inspectors are required to select the annotation they deem correct based on the mechanisms mentioned in \textbf{Section~\ref{sec:pre_n_set}} or provide their own correct annotation if they believe neither of the existing options is accurate. During the music description annotation phase, inspectors independently score each annotation. After the completion of the inspectors' tasks, we collect all the scores for the music descriptions and convene a meeting of the inspectors. We identify data where different inspectors' scores significantly diverge with a maximum difference of more than 10 points, and ask the inspectors to discuss and establish a unified evaluation standard. This training is repeated until the inspectors' scores for the same dataset are roughly consistent.
On the other hand, training for inspectors entails a more intricate process. They must not only master the platform's use but also develop a set of consistent evaluation standards. We gather data annotated by the annotators during the pre-annotation phase and distribute the same dataset to all inspectors. For the lyric annotation phase, inspectors must choose the annotation they consider correct based on the guidelines outlined in \textbf{Section~\ref{sec:pre_n_set}}, or provide an alternative correct annotation if they find the existing ones inaccurate. During the music description annotation phase, inspectors evaluate each annotation independently. Once the inspectors have completed their tasks, we compile all the scores for the music descriptions and organize a meeting with the inspectors. At this meeting, we identify instances where scores from different inspectors significantly vary, with a maximum discrepancy exceeding 10 points, and encourage inspectors to discuss and agree on a unified evaluation criterion. This training process is repeated until the inspectors' scores for the same dataset show substantial consistency.  %TZ

% 接下来我们对标注员和质检员进行相应任务的训练。一方面，标注员需要登录<Untitled>平台，对我们已经聚类好的一小批数据(约20条)进行预标注，包括我们在Section 3.2提到的A, B类的一系列任务。在此期间，标注员需要熟悉平台的操作，明确各个任务应该如何进行标注。我们也针对一些易错问题对标注员进行了针对性培训，例如需要删去原有歌词文本中与歌词本身无关的信息，需要清晰标注出每一个语气词等。（这里可以缩略也可扩展写, 就这样吧不用拓展）
% 另一方面，质检员的训练更为复杂一些：质检员除了需要熟悉平台的操作外，还需要建立统一的评判标准。我们收集了标注员在预标注阶段标注的数据，将这同一批数据分别提供给每一位质检员。在结构标注阶段，质检员需要根据之前提到的机制，对不同标注的数据选择出自己认为正确的标注，或认为二者均不正确，给出自己的正确标注。而在描述标注阶段，质检员需要独立地对每一个标注进行打分。在质检员的工作完成后，我们收集了质检员对所有描述标注任务的打分，并召集质检员会议。我们筛选出不同质检员对同一数据打分出现大幅偏差（最大差值>10分）的数据，要求质检员商议出统一的评判标准。重复上述的训练直到质检员对同一批数据的评分大致相同。

\begin{figure}[htb!]
    \centering
    \includegraphics[width=0.16\textwidth]{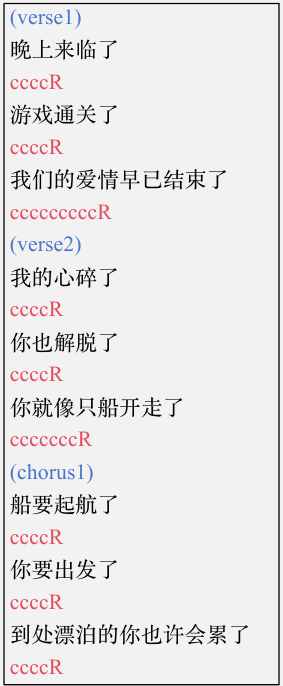}
    \caption{A Fragment from an Illustrative Example of Structure Annotation}
    \label{fig:example_s}
    \vspace{-0.3cm}
\end{figure}

\begin{figure}[htb!]
    \centering
    \includegraphics[width=0.5\textwidth]{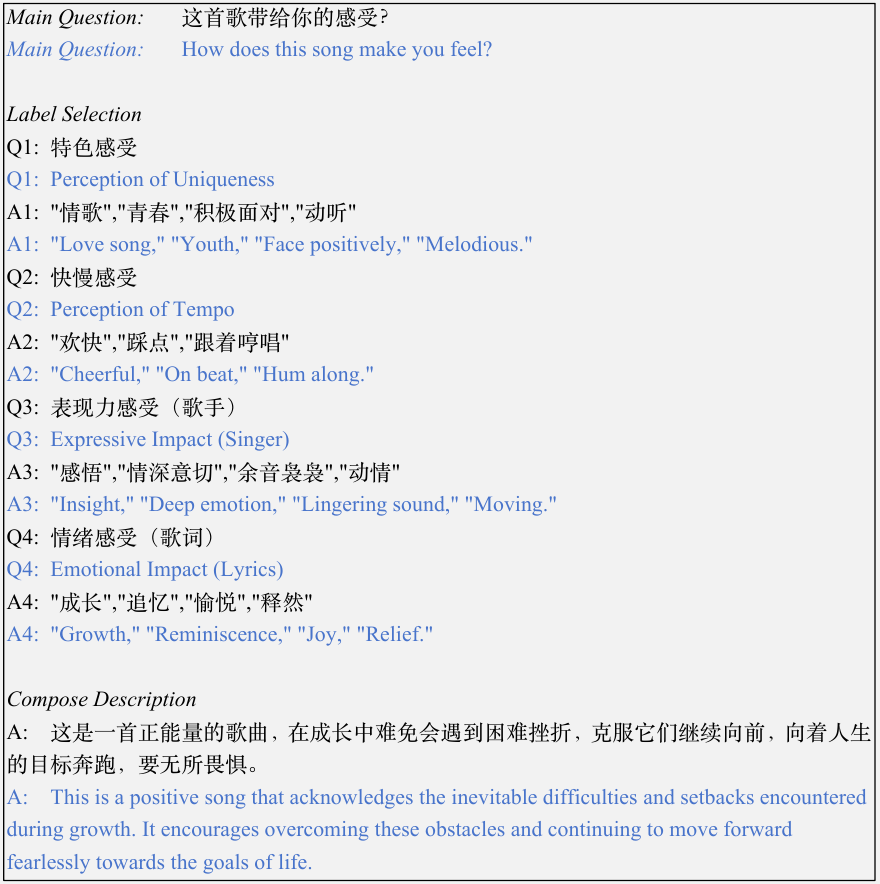}
    \caption{A Fragment from an Illustrative Example of Amateur Description Annotation}
    \label{fig:example_a}
\end{figure}

\begin{figure}[htb!]
    \centering
    \includegraphics[width=0.5\textwidth]{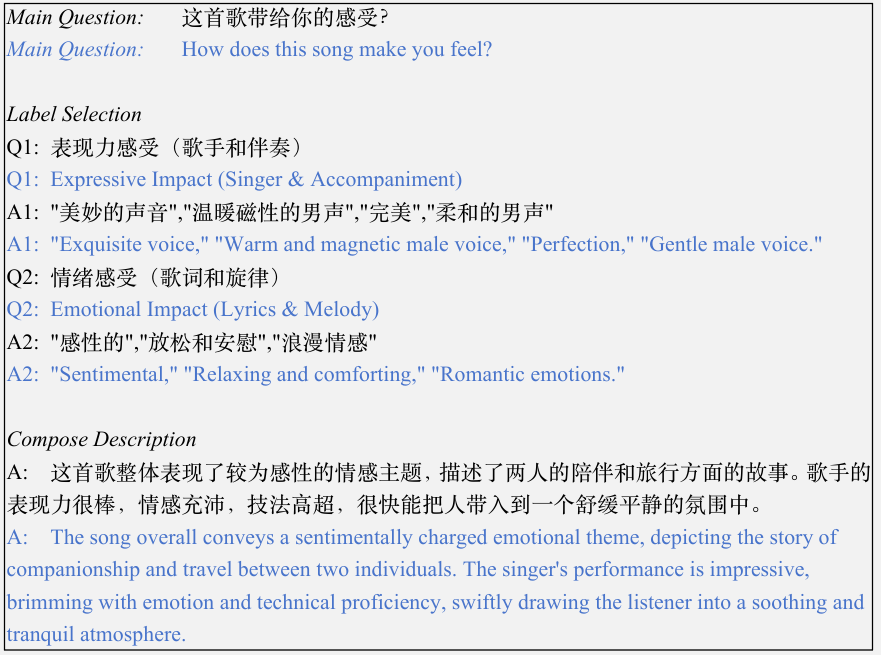}
    \caption{A Fragment from an Illustrative Example of Professional Description Annotation}
    \label{fig:example_p}
\end{figure}

\section{Caichong Music Dataset}
\label{app:caimd}

\subsection{Annotated Data Processing}

%On the one hand, we integrate the annotated information of music sections directly into the lyrics by marking the beginning of each music section with a section label placed before the start of the lyrics for that section. Rhyming information is indicated with strings containing `c' and `R' markers: at the end of each sentence that rhymes with the preceding one, we mark an `R', while non-rhyming parts are marked with a `c'. All the annotated lyric information, including the theme of the lyrics, music sections, and rhyming information, is consolidated in a JSON file using the aforementioned method.
On one hand, we seamlessly incorporate annotations of musical sections into the lyrics by marking the start of each musical section with a section label, positioned before the lyrics of that section begin. We denote rhyming information using strings that include `c' and `R' markers: an `R' is added at the end of any sentence that rhymes with the one before it, while `c' indicates words that do not rhyme. This method is used to compile all annotated lyric information—encompassing the lyrics' theme, musical sections, and rhyming details—into a JSON file.  %TZ

% On the other hand, during the music description annotation phase, we get textual descriptions of each music piece across multiple dimensions, with each annotation including both several descriptive terms and a complete descriptive text. To better enhance the completeness of the descriptions, we concatenate the terms into textual descriptions and merged them into the texts. Subsequently, descriptions from different dimensions are also concatenated, forming a unified and comprehensive annotation that describes the various dimensions of the music.
On the other hand, during the phase dedicated to annotating music descriptions, we collect textual descriptions of each music piece from various perspectives. Each annotation consists of several descriptive terms along with a comprehensive descriptive text. To enhance the richness of these descriptions, we integrate these terms into the textual descriptions, which are then combined with the texts. Furthermore, we concatenate descriptions from different aspects to create a single, detailed annotation that captures the multifaceted nature of the music.  %TZ

% 我们将标注好的乐段信息统一在歌词中，在每个乐段的歌词开始前，以一个乐段标签标记乐段的开始。押韵信息则使用R标记：在每个与先前押韵的句子末尾标记R，不押韵的部分则标记为c。包括歌词主旨、乐段标记和押韵标记在内的结构标注信息被我们用上述方式整合在JSON文件中。

% 在描述标注阶段中，我们将一首歌分多个维度进行了文本描述标注，每条标注包括描述标签和完整文本。为了更好地体现描述的完整性，我们将标签拼接成完整的文本描述，合并入原先的文本描述中。之后，不同维度的文本描述也被拼接，形成一个统一的对音乐各个维度都进行了描述的完整标注。

\subsection{Overview}
\begin{figure}[htb!]
    \centering
    \includegraphics[width=1\linewidth]{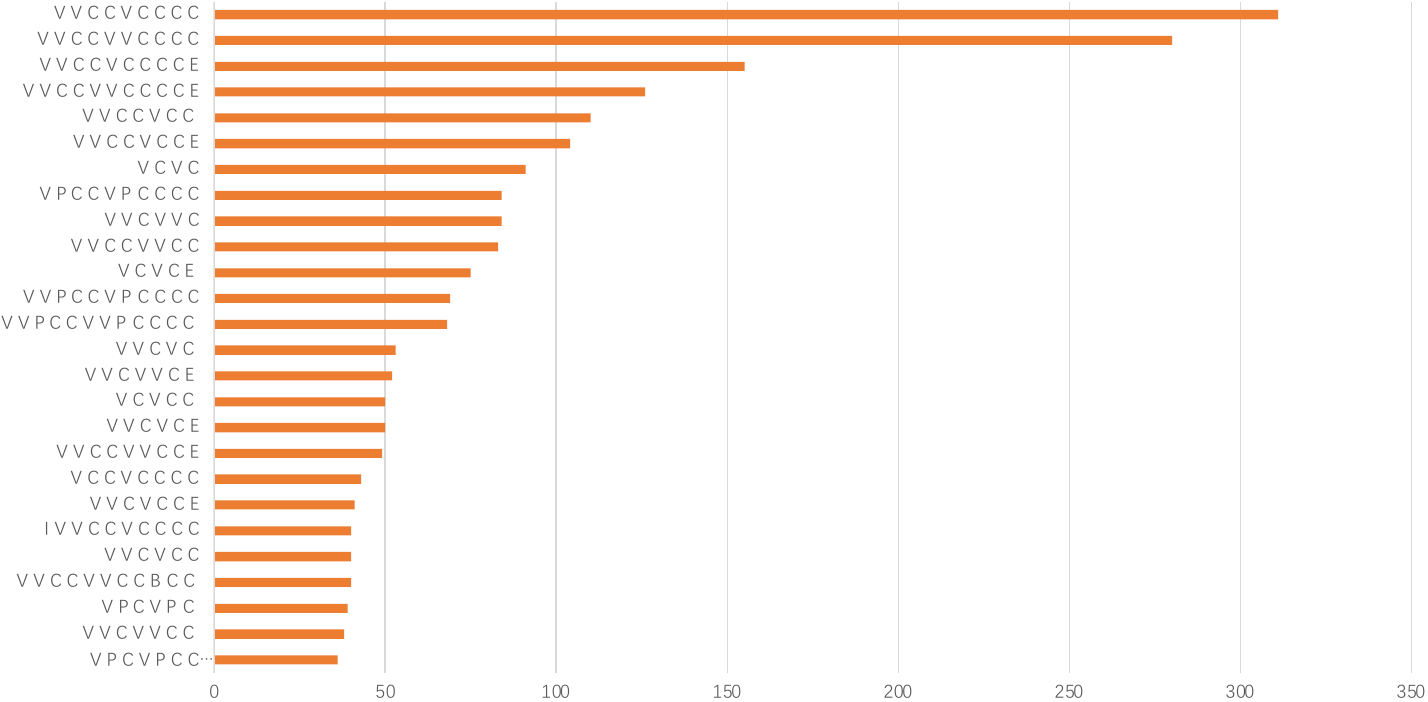}
    \caption{Distribution of Song Structures. The bin labels on the left side of the histogram represent the various musical sections of a song. Specifically, 'i' stands for "Introduction," 'v' corresponds to "Verse," 'c' denotes "Chorus," 'p' indicates "Pre-chorus," 'b' signifies "Bridge," and 'e' represents the "Ending."}
    \label{fig:music_sec}
    \vspace{-0.3cm}
\end{figure}

\begin{figure}[htb!]
    \centering
    \includegraphics[width=1\linewidth]{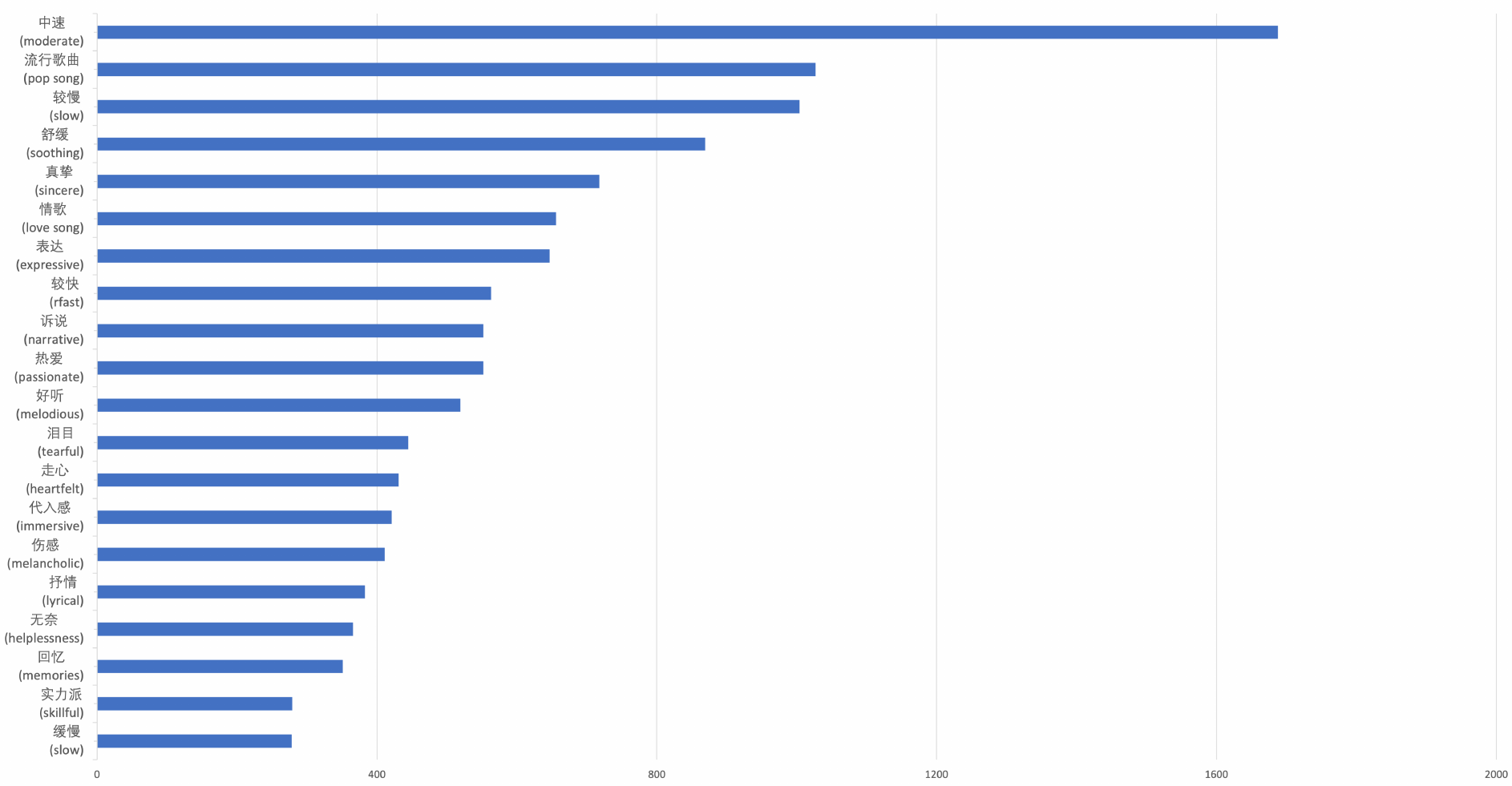}
    \caption{Distribution of Colloquial Descriptive Tags}
    \label{fig:prompts-a}
    \vspace{-0.3cm}
\end{figure}

\begin{figure}[htb!]
    \centering
    \includegraphics[width=1\linewidth]{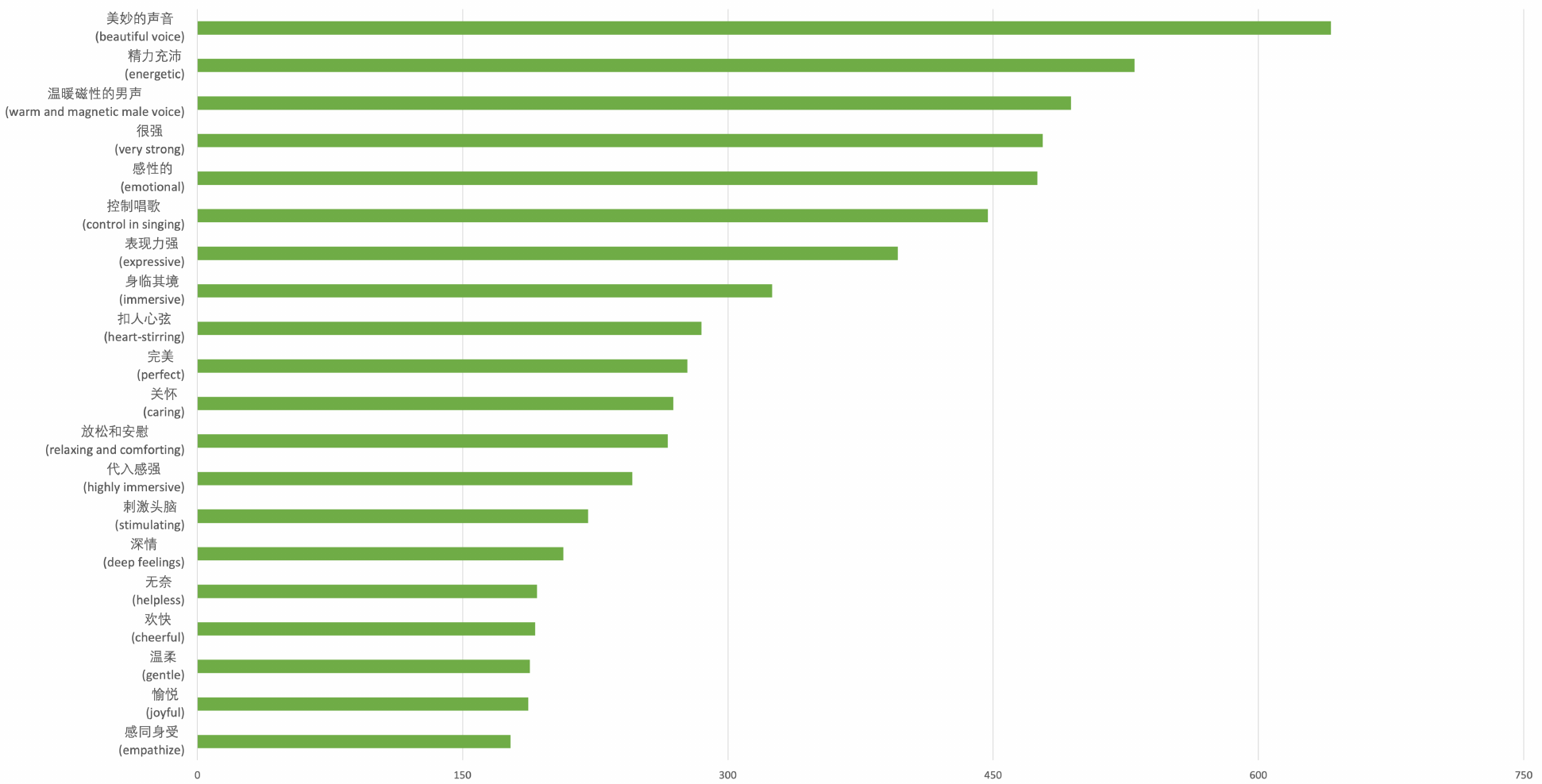}
    \caption{Distribution of Professional Descriptive Tags}
    \label{fig:prompts-p}
    \vspace{-0.3cm}
\end{figure}

% 这里放几张飞书中的overview数据统计分布图
This section provides an overview of the descriptive tag distribution and song structure distribution in CaiMD, as illustrated in Figure~\ref{fig:music_sec}--\ref{fig:prompts-p}. Song structure is the arrangement of musical sections.

\subsection{Examples}
% 这里放一下cccR的样例, 以及标注完成的一首歌的描述. 乐段只放1-2个乐段意思下 不需要放整首歌的结构; 描述也只放1-2个问题意思下,不需要放11个问题,  不用区分专业和业余.

This section presents a range of annotation examples, encompassing both professional and colloquial musical descriptions, along with the musical sections and rhymes featured in CaiMD, as depicted in Figures~\ref{fig:example_s}--\ref{fig:example_p}.

\section{Evaluation Metrics of Structured Lyric Generation}
\label{app:eval_lyr}

\subsection{Formula}

The similarity of the overall structure and musical section structure is calculated according to Equation~\ref{eq:gst1}, where $K_m$ represents the number of matching characters in the longest common subsequence between strings $A$ and $B$. $L_A$ denotes the length of string $A$, and $L_B$ denotes the length of string $B$. In the context of the overall structure, $A$ and $B$ represent the entire set of lyrics. In the context of musical section structure, $A$ and $B$ refer to the sequence of musical section labels.

\begin{equation}
    \centering
    p=\frac{2{{K_m}}}{L_A + L_B}
    \label{eq:gst1}
\end{equation}

The within-section structure similarity is calculated according to Equation~\ref{eq:gst2}. In this equation, each element of $ListA$ and $ListB$ represents the number of sentences contained in each matching musical section of song $A$ and $B$, respectively, e.g., [4, 8, 4] indicates that the three matching musical sections contain 4, 8, and 4 sentences, respectively.

\begin{equation}
    \centering
    p=\frac{2\sum{\min(ListA, ListB)}}{\sum{ListA}+\sum{ListB}}
    \label{eq:gst2}
\end{equation}

Similarly, the within-sentence structure similarity can also be calculated using Equation 2. In this calculation, each element of $ListA$ and $ListB$ represents the number of words in each matching sentence of songs $A$ and $B$.

The calculation of rhyming similarity follows Equation 1, where $K_m$ represents the number of sentences that contain rhyming markers in the lyrics, and $L_A$ and $L_B$ respectively represent the total number of sentences in songs $A$ and $B$.

Since each more detailed structure depends on the match of the preceding structure, cumulative similarity is used when calculating similarity, to take into account the influence of more macroscopic structures on the similarity of more microscopic structures. With the similarities of the overall structure, musical section structure, within-section structure, within-sentence structure, and rhyming structure calculated as $p_1$ to $p_5$ respectively, and their corresponding weights in the overall scoring as $w_1$ to $w_5$, the overall similarity can be calculated using Equation~\ref{eq:overall}.

\begin{equation}
    \centering
    p=\displaystyle\sum_{i=1}^{5}w_i\prod_{j=1}^{i}p_j
    \label{eq:overall}
\end{equation}

Multiplying the overall similarity by 100 gives the overall score. Additionally, the extra reward score based on the proportion of rhyming sentences within the overall lyrics is also incorporated into the overall score.

% 整体结构、乐段结构的相似度按照公式1计算，其中$$K_m$$代表$$A$$$$B$$字符串最大匹配字符长度 (the number of matching characters)。$$L_A$$表示字符串A的长度，$$L_B$$表示字符串B的长度。在整体结构中，$$A$$$$B$$表示全部歌词。在乐段结构中$$A$$$$B$$则表示乐段标签序列。

% 段内结构相似度按照公式2计算。其中ListA, ListB中的每个元素分别表示歌曲A, B每个匹配乐段包含的句子数，e.g. [4, 8, 4]指匹配的3个乐段分别包含4, 8, 4个句子。

% 类似地，句内字数相似度也可以按照公式2计算。在句内字数相似度的计算中，ListA, ListB的每个元素表示歌曲A, B每个相匹配的句子中的字数。
% 押韵相似度的计算按照公式1，$$K_m$$表示歌词包含押韵标记的句子数，$$L_A$$$$L_B$$分别表示歌曲A, B的句子总数。
% 由于每个更加精细的结构都取决于上一个结构的匹配，因此再计算相似度时需要使用累乘相似度，以考虑更宏观的结构对更微观的结构相似度的影响。按照上述方式计算出整体结构、乐段结构、段内结构、句内结构和押韵结构相似度分别为$$p_1$$~$$p_5$$，在整体评分中的权重分别为$$w_1$$~$$w_5$$，则整体相似度可以由公式3计算。

\begin{algorithm}[tb]
    \caption{Reward Score Algorithm}
    \label{alg:reward}
    \textbf{Input}: max\_equ\_slc\_sum, rc\_ing, acmp\_sr, rc\_ino\\
    \textbf{Parameter}: EXTRA\_POINTS\\
    \textbf{Output}: extscore
    \begin{algorithmic}[1] %[1] enables line numbers
        \IF{max\_equ\_slc\_sum == 0}
        \STATE r\_ratio = 0
        \ELSE
        \STATE r\_ratio = rc\_ing / max\_equ\_slc\_sum
        \ENDIF
        \STATE extscore = EXTRA\_POINTS * acmp\_sr
        \IF{0.6 $<=$ r\_ratio \AND r\_ratio $<=$ 0.8}
        \STATE extscore *= 1.0
        \ELSIF{rc\_ino == rc\_ing \AND rc\_ino $>$ 0}
        \STATE extscore *= 0.7
        \ELSE
        \STATE r\_delta = $|$r\_ratio - 0.7$|$
        \IF{r\_delta $<=$ 0.3}
        \STATE extscore *= 0.4 * (1 - r\_delta)
        \ELSE
        \STATE extscore *= 0.0
        \ENDIF
        \ENDIF
        \RETURN extscore
    \end{algorithmic}
\end{algorithm}

\subsection{Reward Score}
% 就是那个额外分的详细计算方法 放个算法这样子

The calculation method of the reward score is shown as Algorithm~\ref{alg:reward}, by which generated lyrics are assigned a certain amount of reward points based on the proportion of rhyming. In this algorithm, max\_equ\_slc\_sum denotes the maximum number of phrases that match; rc\_ing denotes the number of rhyming phrases that match; acmp\_sr denotes the cumulative product of similarities across the first 5 dimensions; rc\_ino denotes the proportion of rhyming within the given rhyme scheme. And EXTRA\_POINTS denotes the total score of the reward score.

\section{Details of Evaluating Music Understanding Models}
\label{app:und}

\subsection{Pipeline of MLP}
% MLP的评测代码架构图可以放一下

\begin{figure}[htb!]
    \centering
    \includegraphics[width=0.8\linewidth]{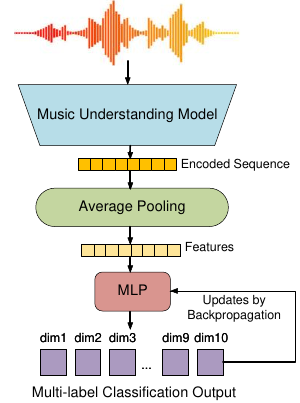}
    \caption{The pipeline of evaluating music understanding models}
    \label{fig:pip-understand}
    \vspace{-0.2cm}
\end{figure}

To assess the effectiveness of music understanding models, we feed music audio into them and obtain their respective encoded sequences. Subsequently, for each model, we utilize an MLP comprising an average pooling layer and 5 linear layers to extract 10 sets of descriptive music tags corresponding to the dimensions of its output encoded sequences. The pipeline of this process can be found in Figure~\ref{fig:pip-understand}.

\newpage

\subsection{Result Analysis}
% 理解模型评估的散点图结果可以放一下

\begin{figure}[htb!]
    \centering
    \includegraphics[width=1\linewidth]{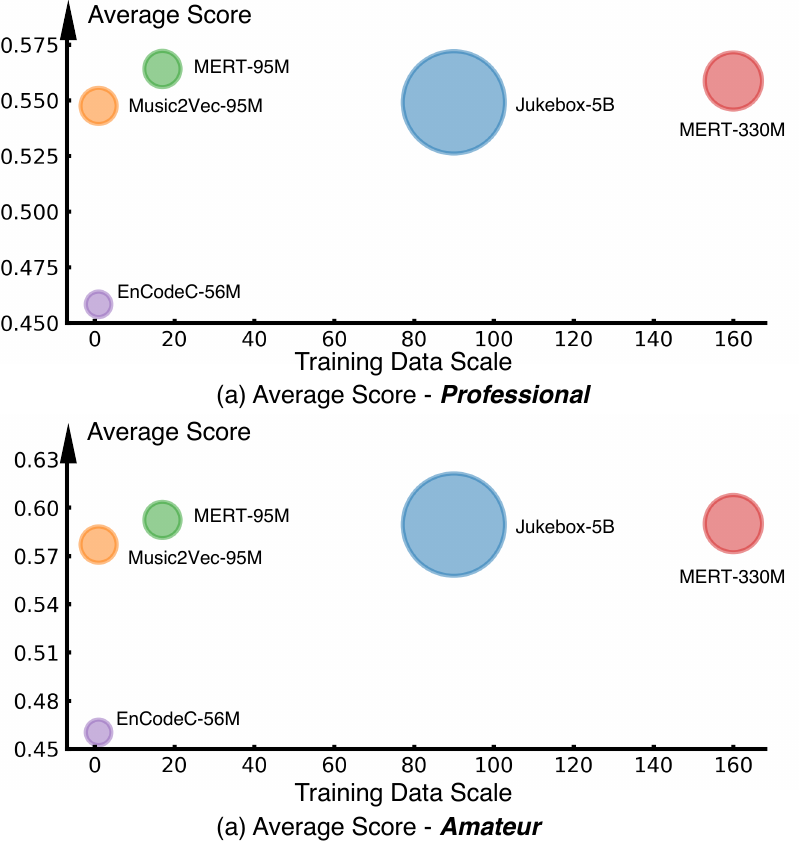}
    \caption{Evaluation of selected music understanding models on the benchmark as represented in a scatter plot.}
    \label{fig:scatter}
    \vspace{-0.3cm}
\end{figure}

Figure \ref{fig:scatter} shows, despite having fewer parameters and a smaller amount of training data, MERT-95M performs best overall in the task of professional and colloquial music description. 

\end{document}